\documentclass{IEEEtaes}

\PassOptionsToPackage{export}{adjustbox}
\usepackage{definitions}
\usepackage[T1]{fontenc}
\usepackage{tikz}
\usepackage{multirow}
\usepackage{flushend}
\usepackage{soul}
\usepackage{siunitx}

\definecolor{mGreen}{rgb}{0.36, 0.69, 0}

\newcommand{\sca}{{\mathscr{a}}}

\newcommand{\uscs}{{\underline{\mathscr{s}}}}
\newcommand{\usca}{{\underline{\mathscr{a}}}}
\newcommand{\usch}{{\underline{\mathscr{h}}}}
\newcommand{\uscb}{{\underline{\mathscr{b}}}}

\newcommand{\sch}{{{\mathscr{h}}}}

\newcommand{\blam}{{\bm{\Lambda}}}

\newcommand{\bmo}{{\bm{\Omega}}}
\newcommand{\uio}{{\underline{\iota}}}
\newcommand{\rg}[1]{{\mathring{#1}}}

\newcommand{\ucb}{{\underline{\mathscr{b}}}}

\newcommand{\acc}{{\ttt{acc}}}
\newcommand{\imu}{{\ttt{imu}}}
\newcommand{\uwb}{{\ttt{uwb}}}

\newcommand{\ttz}{{\ttt{z}}}
\newcommand{\toa}{{\ttt{toa}}}
\newcommand{\tdoa}{{\ttt{tdoa}}}

\newcommand{\tvec}{{{\text{vec}}}}

\newcommand{\hus}{{\hat{\us}}}

\usepackage{etoolbox}
\patchcmd{\thebibliography}{\section*{\refname}}{}{}{}
\pubyear{2025}
\doiinfo{10.1109/TAES.2025.3578410}

\begin{document}
	\title{On Embedding B-Splines in \\Recursive State Estimation}
	
	\author{KAILAI LI}
	\member{Member, IEEE}
	\affil{University of Groningen, Groningen, The Netherlands} 
	
	\receiveddate{Published on IEEE Transactions on Aerospace and Electronic Systems. Personal use of this material is permitted. Permission from IEEE must be obtained for all other uses. DOI: \href{https://doi.org/10.1109/TAES.2025.3578410}{10.1109/TAES.2025.3578410}}
%%    \accepteddate{XXXXX XX XXXX}
%%    \publisheddate{XXXXX XX XXXX}
	
	\corresp{{\itshape (Corresponding author: Kailai Li)}.}
	
	\authoraddress{Kailai Li is with the Bernoulli Institute for Mathematics, Computer Science and Artificial Intelligence, University of Groningen, 9747 AG Groningen, The Netherlands (e-mail: \texttt{\href{mailto:kailai.li@rug.nl}{kailai.li@rug.nl}})}
	
	\supplementary{Color versions of one or more of the figures in this article are available online at \href{http://ieeexplore.ieee.org}{http://ieeexplore.ieee.org}.}
	
	\markboth{KAILAI LI}{ON EMBEDDING B-SPLINES IN RECURSIVE STATE ESTIMATION}
	
	\maketitle
	\begin{abstract}
		We present a principled study on establishing a recursive Bayesian estimation scheme using B-splines in Euclidean spaces. The use of recurrent control points as the state vector is first conceptualized in a recursive setting. This enables the embedding of B-splines into the state-space model as a continuous-time intermediate, bridging discrete-time state transition with asynchronous multisensor observations. Building on this spline-state-space model, we propose the spline-embedded recursive estimation scheme for general multisensor state estimation tasks. Extensive evaluations are conducted on motion tracking in sensor networks with time-difference-of-arrival and time-of-arrival-inertial settings using real-world and real-world-based synthetic datasets, respectively. Numerical results evidently demonstrate several advantages of spline embedding in recursive state estimation over classical discrete-time filtering approaches in terms of tracking accuracy, robustness, and memory efficiency.
	\end{abstract}
	
	\begin{IEEEkeywords}
		Estimation, filtering, stochastic systems
	\end{IEEEkeywords}
	%--------------------------------------------------------------------------------------
	
	\section{\uppercase{Introduction}}\label{sec:introduction}
	State estimation is fundamental to the development of reliable mobile applications across indoor, ground, ocean, and aerospace scenarios~\cite{kassas2024ad,inalhan2002relative,dalberg2006underwater,wang2021challenges,kok2015indoor}, underpinning essential functions such as target tracking, localization, navigation, guidance, and control~\cite{gustafsson2010particle,bar1990tracking,willett2002pmht,skog2010zero,gustafsson2005mobile}. In many engineering practices, performing state estimation involves handling asynchronous measurements from multiple sensors, with common modalities covering sonar, radar, camera, ultra-wideband (UWB), light detection and ranging (LiDAR), and inertial measurement unit (IMU)~\cite{engel12vicomor,ECC20_Li,xu2022fast,kenney2023cooperative,RAL23_Li,ma2024long,ICRA20_Li}.
	
	Classical approaches to multisensor motion tracking are fundamentally rooted in the methodology of recursive Bayesian estimation. Common techniques include the extended Kalman filter (EKF), the unscented Kalman filter (UKF), and the particle filter~\cite{bloesch2017rovio,julier2004unscented,fredrik2002,LCSS21_Li}. These methods rely on state-space representation to model uncertain physical systems in discrete time, incorporating both process and measurement models under uncertainty~\cite{kay1993fundamentals}. Based thereon, state estimates are recursively computed through prediction and update steps over time, with built-in probabilistic uncertainty quantification. This enables downstream functions and tasks, such as outlier rejection, planning, decision-making, and control, to be performed with sufficient accuracy and confidence~\cite{li2021tsp,chen2011kalman,Fusion19_Bultmann}. To estimate the 6-DoF rigid-body motions, recursive estimation approaches have been developed w.r.t.\ nonlinear manifolds using error-state formulation and geometry-driven methodologies~\cite{sola2017quaternion,Li2022Dissertation}.
	
	The conventional recursive estimation scheme relies on the Markov assumption, which ideally presumes prior knowledge of the underlying system dynamics and restricts state propagation to consecutive time steps. Recent advances in mobile sensing have popularized alternative approaches based on nonlinear optimization (smoothing). In online tracking tasks, motion states are typically estimated in a sliding-window fashion via maximum likelihood estimation (MLE) or maximum a posteriori (MAP), with residual terms incorporating multisensor measurements formulated over discrete-time factor graphs~\cite{slam2010}. Compared to recursive methods, this paradigm shift has led to improved estimation accuracy while requiring greater, yet still tractable, computational resources, by enabling sparse structure of the optimization problem~\cite{strasdat2010why}. In adaptation to multisensor fusion, a variety of key techniques, such as IMU preintegration and keyframe-based sliding-window optimization, have been proposed to achieve high-performance state estimation in terms of accuracy, robustness and runtime efficiency~\cite{vins2018,yan2024plpf,RAL21_Li}.
	
	Methodologically speaking, the evolution of system states is governed by underlying stochastic dynamics that are inherently continuous over time, while observations are made at discrete timestamps through specific sensor modalities. To properly address this discrepancy, most recursive estimation schemes rely on temporal discretization of the continuous-time state-space models, resulting in the continuous-discrete and discrete-time variants~\cite{sarkka2013bayesian}. For that, remedies such as oversampling of the system or approximation of transition densities have been introduced, and additional efforts are typically required for achieving viable accuracy and stability~\cite{axe2015dis}. Given asynchronous sensor readings, delivering state estimates at the desired time instants (e.g., at a fixed frame rate) inevitably requires certain approximation, such as linear interpolation~\cite{vins2018,RAL21_Li}. Besides, sensor measurements that contribute to the state propagation are often assumed to be constant over the sampling interval, introducing additional errors. These operations often require tedious implementation, especially when combined with other preprocessing procedures, e.g., IMU preintegration~\cite{vins2018} or motion undistortion of LiDAR scans~\cite{xu2022fast}. Furthermore, enforcing kinematic relations over sequential motion estimates in conjunction with sensor fusion is hardly possible without introducing explicit constraints. Under unfavorable conditions, such as highly dynamic motions and noisy measurements with outliers, state estimates can exhibit physically infeasible transitions~\cite{RAL23_Li}.
	
	There have been growing interest in enabling continuous-time sensor fusion, with significant progress made in the use of B-splines~\cite{anderson2013spline,mueggler15rss}. Established atop a series of control points associated with temporal knots, B-splines parameterize motion trajectories as polynomial functions over time, leading to a more efficient state representation in comparison with the discrete-time counterpart~\cite{cio2022spline}. B-splines are smooth to a given order and inherently guarantee kinematic relations over temporal differentiations. Additionally, the locality of B-splines enables convenient interpolations via matrix multiplication and implicitly spreads correlations over dynamical motions through adjacent control points~\cite{qin1998general}. Using the cumulative form of B-splines, it is also possible to extend their utility from modeling vector-valued (Euclidean) to Lie group-valued functions. This makes them appealing for continuous-time parameterization of 6-DoF pose trajectories in navigation scenarios~\cite{furgale2012spline,sommer2016continuous}.
	
	Existing B-spline-based motion estimation approaches uniformly rely on nonlinear optimization, typically formulated as MLE or MAP and solved in a batch-wise or sliding-window fashion. This scheme has been adopted in a few multisensor localization and navigation systems, involving modalities such as IMU, (event) camera, radar, UWB, and LiDAR~\cite{mueggler2018continuous,hug2022con,cio2022spline,RAL23_Li,lv2021clins,yang2021spline}. Here, B-spline control points are optimally estimated via nonlinear least squares, with residuals directly comparing sensor measurements to their interpolated values at exact observation timestamps without complex preprocessing. To enable high-performance sensor fusion, several key theoretical and practical advancements have been introduced, such as quaternion-based B-splines for 6-DoF motion modeling, closed-form temporal and spatial differentiations, and customized nonlinear solvers for spline fitting~\cite{sommer2020efficient,RAL23_Li}.
	
	While the recursive filtering scheme offers lightweight and probabilistic computational framework, its integration with the B-spline state modeling remains largely unexplored, particularly in the pursuit of developing a self-contained state estimation methodology. The only existing relevance is confined to the application of recursive filters to geometric tasks, such as spline-based curve or surface modeling, without any connection to state estimation theory or multisensor engineering practice~\cite{splinekf1997,jauch201728}. To the best knowledge of the author, there has been no systematic study into exploiting B-splines in recursive state estimation for uncertain dynamical systems.
	
	\subsection*{Contributions}
	In light of the state of the art, we propose a recursive Bayesian estimation framework for continuous-time motion estimation in multisensor settings. B-splines are employed for Euclidean state parameterization, enabling unified kinematic interpolations through a concise matrix-based formulation and expressive modeling of underlying uncertain dynamics. We conceptualize the use of recurrent control points as the state vector and introduce the spline-state-space (TriS) model based thereon, thereby endowing B-spline-based state estimation with a principled probabilistic interpretation. Building on this foundation, we develop the spline-embedded recursive estimation (SERE) scheme and investigate its fundamental probabilistic properties. The proposed scheme is validated through extensive experiments on both real-world and real-world-based synthetic datasets for motion tracking in sensor networks, using time-difference-of-arrival (TDoA) and time-of-arrival (ToA)-inertial configurations, respectively. Compared to conventional discrete-time filtering methods, the proposed B-spline embedding significantly improves tracking accuracy, robustness, and memory efficiency. As the first fundamental study of spline-based recursive state estimation, this work sketches out a self-contained paradigm for continuous-time probabilistic multisensor state estimation.
	
	The remainder of this article is organized as follows. Notation conventions are introduced in \secref{sec:notation}, followed by the fundamentals of B-spline motion parameterization in \secref{sec:spline}. The proposed TriS model, including the conceptualization of recurrent control points, is presented in \secref{sec:tris}. \secref{sec:filter} will further elaborate the proposed spline embedding for state estimation. The extensive evaluations on multisensor motion tracking will be provided in \secref{sec:eva}, and the paper is concluded in \secref{sec:conc}.
	
	\section{\uppercase{Notation Conventions}}\label{sec:notation}
	Throughout this article, scalar quantities are denoted by lowercase letters, e.g., $x\in\R$, and vectors are indicated by lowercase letters with an underline, such as $\ux\in\R^d$, where $d$ specifies the dimension of the Euclidean space. Matrices are represented by bold uppercase letters, such as $\fD\in\R^{d\times{d}}$. Specifically, we use $\fI_d\in\R^{d\times{d}}$ and $\fzero_d\in\R^{d\times{d}}$ to denote $d$-dimensional identity and zero matrices, respectively. In the context of recursive state estimation, the discrete time step is denoted by $k\in\N$, whereas the exact timestamp at step $k$ is written as $t_k$. Additionally, calligraphic lowercase letters represent functions, such as $\sca(x)$ and $\usca(x)$ for scalar- and vector-valued functions, respectively. The operator $\otimes$ denotes the Kronecker product, and $\oplus$ represents the direct sum of two matrices~\cite{rosen1999handbook}.
	
	\begin{figure*}[t]
		\centering
		\includegraphics[width=0.98\textwidth]{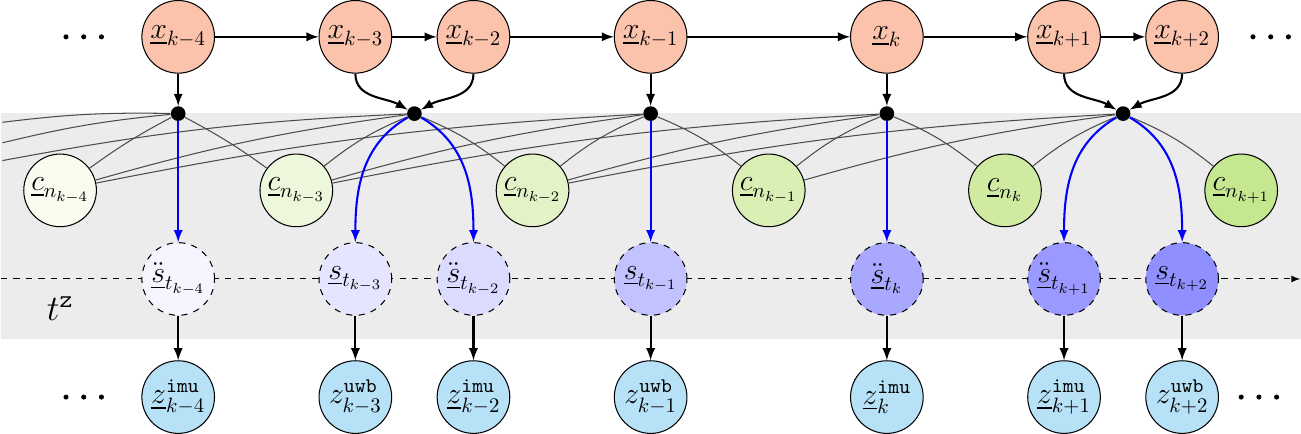}
		\caption{Graph representation of the proposed spline-state-space (TriS) model. A cubic B-spline is embedded into the state-space model as a continuous-time intermediate (indicated by the gray area), bridging discrete-time state transitions and sensor observations. Nodes representing recurrent control points (lime green) within the state vectors (red) and interpolated dynamical motions (purple) queried by the measurements (blue) are depicted processing from lighter to darker colors over time. We showcase asynchronous ultra-wideband and inertial measurements (denoted by superscripts $\uwb$ and $\imu$, respectively) corresponding to the interpolated position ($\us$) and acceleration ($\ddot{\us}$), respectively.}
		\label{fig:tris}
	\end{figure*}
	
	\section{\uppercase{Continuous-Time Motion Representation}}\label{sec:spline}
	For demonstrating our work in this paper, we express B-splines using their cubic form (fourth-order) in $d$-dimensional Euclidean spaces. Note that the presented techniques can be extended to B-splines of higher orders in a straightforward manner. Cubic B-splines exhibit continuous-time derivatives up to the second order, which is sufficient for fusing measurements of most sensor modalities related to motion estimation, including the accelerometer~\cite{RAL23_Li,lv2021clins,yang2021spline}. 
	
	Given a set of control points $\{\uc_i\}_{i=1}^n\subset\R^d$ associated with equidistant temporal knots $\{t_i\}_{i=1}^n\subset\R$ of temporal interval $\tau$, a cubic B-spline can be established for trajectory representation. At an arbitrary timestamp $t\in(\,t_i,t_{i+1}]$, the corresponding trajectory point is determined by a local set of control points $\{\uc_{i+j-2}\}_{j=0}^3$ according to
		\begin{equation}\label{eq:kino0}
			\uscs(t)=\fC_i\,\bmo\,\uu_t\in\R^d\,,
		\end{equation}
	with matrix 
	\begin{equation*}
		\fC_i=[\,\uc_{i-2},\,\uc_{i-1},\,\uc_{i},\,\uc_{i+1}\,]\in\R^{d\times4}
	\end{equation*}
	concatenating the corresponding control points columnwise. The vector $\uu_t$ is given by
	\begin{equation*}
		\uu_t=[\,1,u_t,u_t^2,u_t^3\,]^\top\in\R^4\,,
	\end{equation*}
	which contains the powers of the normalized time 
	\begin{equation}\label{eq:u}
		u_t=(t-t_i)/\tau\,.
	\end{equation}
	$\bmo$ in \eqref{eq:kino0} is the \textit{basis function matrix} of cubic B-splines and follows
	\begin{equation}\label{eq:itpMatcompact}
		\bmo=\frac{1}{6}\bbmat\,1&-3&3&-1\,\\4&0&-6&3\\1&3&3&-3\\0&0&0&1\ebmat\,.
	\end{equation}
	We can derive the velocity trajectory of cubic B-splines by taking the first derivative of the polynomial function \eqref{eq:kino0} \wrt time. It is given by
	\begin{equation}\label{eq:kino1}
		\dot{\uscs}(t)=\fC_i\,\bmo\,\dot{\uu}_t\,,\eqwith \dot{\uu}_t=[\,0,1,2u_t,3u_t^2\,]^\top/\tau\,.
	\end{equation}
	Further, the acceleration trajectory can be derived as the following function over time
	\begin{equation}\label{eq:kino2}
		\ddot{\uscs}(t)=\fC_i\,\bmo\,\ddot{\uu}_t\,,\eqwith\ddot{\uu}_t=[\,0,0,2,6u_t\,]^\top/\tau^2\,.
	\end{equation}
	In the following content, we use $\circ$ atop the spline function $\uscs(t)$, i.e., $\rg{\uscs}(t)$, to denote trajectories of different orders. This includes position ${\uscs}(t)$, velocity $\dot{\uscs}(t)$, and acceleration $\ddot{\uscs}(t)$, which can be unified through the following expression for kinematic interpolation
	\begin{equation}\label{eq:kino}
		\rg{\uscs}(t)=\fC_i\,\bmo\,\rg{\uu}_t\in\R^d\,,
	\end{equation}
	where $\rg{\uu}_t\in\R^4$ are explicitly detailed in \eqref{eq:kino0}, \eqref{eq:kino1}, and \eqref{eq:kino2}.
	
	\section{\uppercase{Stochastic Modeling on B-Splines}}\label{sec:tris}
	In existing multisensor B-spline fitting methods, control points over a certain time span are typically estimated through nonlinear least squares optimization. The kinematic interpolation in \secref{sec:spline} enables residuals to directly incorporate sensor measurements at their precise observation timestamps~\cite{sommer2020efficient,RAL23_Li}. However, modeling the uncertainty of the resulting motion estimates remains challenging within the optimization-based scheme. To address this limitation, we introduced a principled framework--the spline-state-space (TriS) model--for continuous-time stochastic modeling based on B-splines.
	
	\subsection{Vectorized Kinematic Interpolations}\label{subsec:reform}
	We now reformulate the basic kinematic interpolations introduced in \secref{sec:spline} using a vectorized representation of the set of control points. According to~\cite{macedo2013typing}, we perform vectorization on both sides of \eqref{eq:kino} and obtain
	\begin{equation*}
		\tvec(\rg{\uscs}(t))=\tvec(\fC_i\,\bmo\,\rg{\uu}_t)=((\bmo\,\rg{\uu}_t)^\top\otimes\fI_d)\,\tvec(\fC_i)\,,
	\end{equation*}
	with $t\in(t_i,t_{i+1}\,]$\,. $\otimes$ is the Kronecker product. Note that vectorizing $\rg{\uscs}(t)\in\R^d$ does not alter its value. We obtain
	\begin{equation}\label{eq:kinoVec}
		\rg{\uscs}(t)=\rg{\blam}_t\,\ux_i\,,\eqwith \rg{\blam}_t=(\bmo\rg{\uu}_t)^\top\otimes\fI_d\in\R^{d\times4d}
	\end{equation}
	containing the interpolation coefficients \wrt control points concatenated into the vector
	\begin{equation*}
		\ux_i=\tvec(\fC_i)=[\,\uc_{i-2}^\top,\,\uc_{i-1}^\top,\,\uc_{i}^\top,\,\uc_{i+1}^\top\,]^\top\in\R^{4d}.
	\end{equation*}
	
	\subsection{Spline-State-Space (TriS) Model}\label{subsec:tris}
	In the context of recursive Bayesian estimation, B-splines can be leveraged for continuous-time parameterization of dynamical motions in adaptation to discrete-time observations. We conceptualize this idea into the so-called spline-state-space (TriS) model as illustrated by the graph representation in \figref{fig:tris}. A B-spline is embedded to a typical state-space model as an intermediate between the discrete-time states and observations. 
	
	\subsubsection*{State vector} 
	Given the measurements $\{\uz_i\}_{i=1}^k$, the cubic B-spline is maintained using a minimal set of control points $\{\uc_{i}\}_{i=1}^{n_k}$, placed at timestamps $\{t_i\}_{i=1}^{n_k}$ with a uniform knot interval $\tau$. The latest measurement $\uz_k$ falls within the final knot interval, specifically $t_k^\ttz\in(\,t_{n_k-1},t_{n_k}]$, where the corresponding B-spline segment is governed by the most recent four control points $\{\uc_{n_k-i}\}_{i=0}^3$. These are composed into the following state vector
	\begin{equation}\label{eq:state}
		\ux_k=[\,\uc_{n_k-3}^\top,\uc_{n_k-2}^\top,\uc_{n_k-1}^\top,\uc_{n_k}^\top\,]^\top\in\R^{4d}\,.
	\end{equation}
	We conceptualize this local set of control points as the \textit{recurrent control points} (RCPs), emphasizing their temporal alignment with the most recent measurement as it evolves over time.
	
	\subsubsection*{Measurement model} 
	Given the state vector in \eqref{eq:state}, the measurement model at step $k$ can be expressed as the following general form
	\begin{equation}\label{eq:meas}
		\uz_{k}=\usch_k(\ux_{k})+\uv_k\,,\quad\text{where}\quad t_k^\ttz\in(\,t_{n_k-1},t_{n_k}]
	\end{equation}
	denotes the associated observation timestamp. We use $\Z$ to represent the domain of measurements. $\uv_k\in\Z$ denotes a zero-mean additive noise term of covariance $\fR_{k}$. The observation function $\usch_k\colon\R^{4d}\rightarrow\Z$ maps the RCPs to the measurement space at observation timestamp $t_k^\ttz$. It can be decomposed into two cascaded steps as follows
	\begin{equation}\label{eq:fh}
		\usch_{k}(\ux_k)=\ucb\big(\rg{\uscs}(t_k^\ttz;\ux_k)\big)\,.
	\end{equation}
	$\rg{\uscs}(t_k^\ttz;\ux_k)$ is the spline function determined by the RCPs in $\ux_k$ evaluated at timestamp $t_k^\ttz$ according to \eqref{eq:kinoVec}. It generates the motion variable for sensing. Further, function $\uscb:\R^d\rightarrow\Z$ models the actual sensor modality, mapping the interpolated motion to the noise-free observation. In \figref{fig:tris}, this cascaded measurement modeling is showcased using an ultra-wideband-accelerometer configuration.
	
	\subsubsection*{Process model} The process model describes the propagation of recurrent control points in the case that the incoming measurement falls outside the current spline time span $[\,t_1,t_{n_k}]$ over the $n_k$ knots. Suppose a new measurement $\uz_{k+1}$ arrives at timestamp $t_{k+1}^\ttz\in(\,t_{n_k},t_{n_k}+\tau\,]$. To formulate the measurement model at $t_{k+1}^\ttz$ as per \eqref{eq:meas}, the spline must be extended to span $n_{k+1}=n_k+1$ knots. Consequently, the state vector defined in \eqref{eq:state} is updated to
	\begin{equation}\label{eq:stateNew}
		\begin{aligned}
			\ux_{k+1}&=[\,\uc_{n_{k+1}-3}^\top,\uc_{n_{k+1}-2}^\top,\uc_{n_{k+1}-1}^\top,\uc_{n_{k+1}}^\top\,]^\top\\
			&=[\,\uc_{n_{k}-2}^\top,\uc_{n_{k}-1}^\top,\uc_{n_k}^\top,\uc_{n_{k}+1}^\top\,]^\top\,,
		\end{aligned}
	\end{equation}
	where the first three control points overlap with the last three control points in the previous state $\ux_k$ in \eqref{eq:state}. The general process model is expressed as
	\begin{equation}\label{eq:sys}
		\ux_{k+1} = \usca(\ux_k)+\uw_{k}\,,
	\end{equation}
	with $\uw_{k}\in\R^{4d}$ being the additive process noise assumed to be zero-mean with covariance $\fQ_{k}\in\R^{4d\times4d}$. The transition function $\usca\colon\R^{4d}\rightarrow\R^{4d}$ in \eqref{eq:sys} can be established with reference to various dynamical principles. We hereby introduce a straightforward strategy with the following two steps: 1) we retain the first three control points $\{\uc_{n_k-i}\}_{i=0}^2$ in $\ux_{k+1}$ at their coordinates in the previous state $\ux_k$, and 2) the latest control point $\uc_{n_k+1}$ is added by preserving the velocity at timestamp $t_{n_k-1}$. For the latter step, we perform velocity interpolations at timestamps $t_{n_k-1}$ and $t_{n_k+1}$ according to \eqref{eq:kino1}, resulting in
	\begin{equation*}
		\begin{aligned}
			\dot{\uscs}({t_{n_k-1}})&=\frac{\uc_{n_k-1}-\uc_{n_k-3}}{2\tau}\eqand\\
			\dot{\uscs}({t_{n_k+1}})&=\frac{\uc_{n_k+1}-\uc_{n_k-1}}{2\tau}\,,
		\end{aligned}
	\end{equation*}
	respectively. Imposing identical velocities at these two timestamps then yields 
	\begin{equation*}
		\uc_{n_k+1}=2\uc_{n_k-1}-\uc_{n_k-3}\,.
	\end{equation*}
	Therefore, the process model in \eqref{eq:sys} can be concretized as the following linear expression
	\begin{equation}\label{eq:sysCV}
		\ux_{k+1}=\fA\,\ux_k+\uw_{k}\,,
	\end{equation}
	with the transition matrix given by
	\begin{equation*}
		\fA=\bbmat \fzero_d &\fI_d&\fzero_d&\fzero_d&\\
		\fzero_d &\fzero_d &\fI_d &\fzero_d\\\fzero_d&\fzero_d&\fzero_d&\fI_d\\
		-\fI_d &\fzero_d &2\fI_d &\fzero_d\ebmat.
	\end{equation*}
	\begin{figure*}[t]
		\centering
		\begin{tabular}{cccc}
			\adjustbox{trim={0.03\width} {0.01\height} {0.09\width} {0.07\height},clip}{\includegraphics[width=0.255\textwidth]{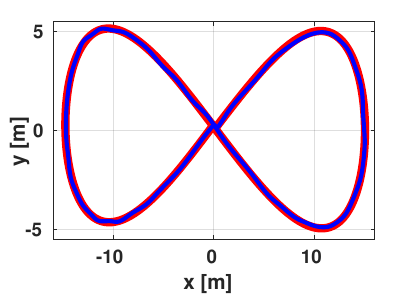}} &
			\adjustbox{trim={0.03\width} {0.01\height} {0.09\width} {0.07\height},clip}{\includegraphics[width=0.255\textwidth]{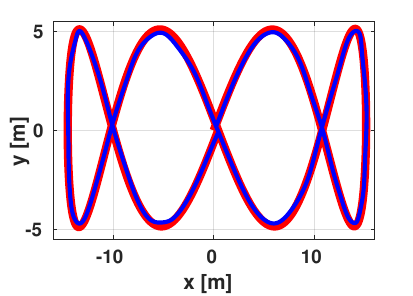}} &
			\adjustbox{trim={0.03\width} {0.01\height} {0.09\width} {0.07\height},clip}{\includegraphics[width=0.255\textwidth]{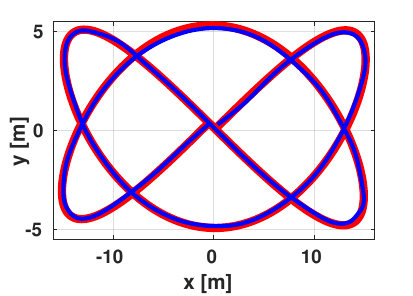}} &
			\adjustbox{trim={0.03\width} {0.01\height} {0.09\width} {0.07\height},clip}{\includegraphics[width=0.255\textwidth]{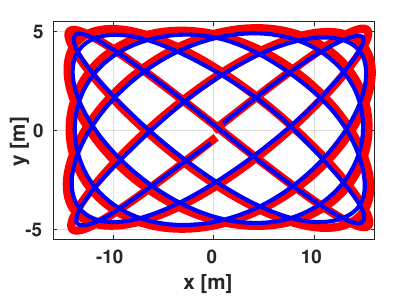}} \\
			(A) &(B) &(C) &(D)
		\end{tabular}
		\caption{Exemplary runs of the proposed spline-embedded recursive estimation for tracking along various Lissajous curves. Blue and red trajectories indicate the estimates and the ground truth, respectively.}
		\label{fig:lissa}
	\end{figure*}
	\begin{figure*}[t]
		\centering
		\begin{tabular}{cccc}
			\adjustbox{trim={0.08\width} {0.01\height} {0.09\width} {0.01\height},clip}{\includegraphics[width=0.57\textwidth]{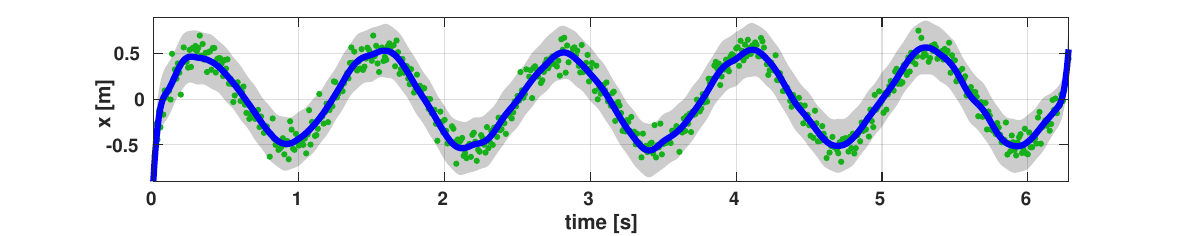}} &
			\adjustbox{trim={0.08\width} {0.01\height} {0.09\width} {0.01\height},clip}{\includegraphics[width=0.57\textwidth]{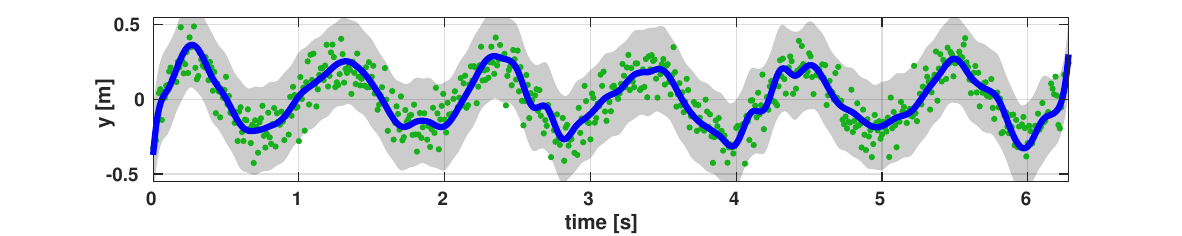}}
		\end{tabular}
		\caption{Estimates on the $x$ and $y$ axes over the sequence drawn in \figref{fig:lissa}-(D). The means are depicted in blue \wrt the ground truth, and the estimated uncertainty is plotted in gray using two standard deviations, including measurement noise. Green dots indicate the position measurements w.r.t.\, the estimates.} 
		\label{fig:lissasig}
	\end{figure*}
	
	\section{\uppercase{Spline Embedding in Recursive Estimation}}\label{sec:filter}
	Based upon the proposed spline-state-space model, the task of continuous-time motion estimation can be converted into estimating discrete-time recurrent control points. To operationalize this methodology, we now introduce the spline-embedded recursive estimation (SERE) scheme, including a case study on motion tracking presented in this section.
	
	\begin{algorithm}[t]
		\caption{ \strut{Spline-Embedded Recursive Estimation}}\label{alg:filter}
		\KwIn{posterior estimate $\hux_{k-1\vert{k-1}},\fP_{k-1\vert{k-1}}$ measurement $\uz_k$ at $t_k^\ttz$, knot quantity ${n_{k-1}}$}
		\KwOut{posterior estimate $\hux_{k\vert{k}}, \fP_{k\vert{k}}$}
		\If{$t_k^\ttz>t_{n_{k-1}}$}{
			\tcc{prediction}
			$n_k\leftarrow n_{k-1}+1$\,\;
			$\hux_{k\vert{k-1}}\leftarrow\fA\hux_{k-1\vert{k-1}}$\,\;
			$\fP_{k\vert{k-1}}\leftarrow\fA\fP_{k-1\vert{k-1}}\fA^\top+\fQ_k$\,\;
		}\Else{
			$n_{k}\leftarrow n_{k-1}$\,\;
			$\hux_{k\vert{k-1}}\leftarrow\hux_{k-1\vert{k-1}}$\,\;
			$\fP_{k\vert{k-1}}\leftarrow\fP_{k-1\vert{k-1}}$\,\;
		}
		\tcc{update}
		$\fH_k\leftarrow\fJ_{\ucb}(\rg{\uscs}(t_k^\ttz;\hux_{k\vert{k-1}}))\rg{\blam}_{t_k^\ttz}$\,;\quad\tcp{see \eqref{eq:hmat}}
		$\fK_k\leftarrow\fP_{k\vert{k-1}}\fH_k^\top(\fH_k\fP_{k\vert{k-1}}\fH_k^\top+\fR_k)^{-1}$\,\;
		$\hux_{k\vert{k}}\leftarrow\hux_{k\vert{k-1}}+\fK_k(\uz_k-\usch_k(\hux_{k\vert{k-1}}))$\,\;
		$\fP_{k\vert{k}}\leftarrow(\fI-\fK_k\fH_k)\fP_{k\vert{k-1}}$\,\;
		\Return $\hux_{k\vert{k}}\,,\fP_{k\vert{k}}$
	\end{algorithm}
	
	\subsection{Spline-Embedded Recursive Estimation (SERE)}\label{subsec:sere}
	As shown in \algref{alg:filter}, the recurrent control points are estimated recursively following the concept of Kalman filtering based on the TriS model. Suppose that a measurement $\uz_k$ is obtained at timestamp $t_k^\ttz$, given $n_{k-1}$ control points. The state $\ux_{k-1}$
	composes recurrent control points $\{\uc_{n_{k-1}-i}\}_{i=0}^3$, and its previous posterior estimate mean $\hux_{k-1\vert{k}-1}$ and covariance $\fP_{k-1\vert{k}-1}$ are available. If the measurement timestamp exceeds the current spline time span, namely, $t_k^\ttz>t_{n_{k-1}}$, we perform state propagation according to the system  
	model in \eqref{eq:sysCV}, leading to $n_{k}=n_{k-1}+1$ control points and the predicted prior estimate (\algref{alg:filter}, lines 1--4). Otherwise, the previous state estimate remains as the prior, with the number of control points being unchanged (\algref{alg:filter}, lines 5--8). During the update step, a basic strategy is to linearize the observation function \eqref{eq:fh} in the measurement model \eqref{eq:meas} at the prior estimate mean $\hux_{k\vert{k-1}}$. The resulting observation matrix can be obtained by applying the chain rule as
	\begin{equation}\label{eq:hmat}
		\begin{aligned}
			\fH_k&=\fJ_{\ucb}(\rg{\uscs}(t_k^\ttz;\hux_{k\vert{k-1}}))\,\fJ_{\rg{\uscs}}(t_k^\ttz;\hux_{k\vert{k-1}})\\
			&=\fJ_{\ucb}(\rg{\uscs}(t_k^\ttz;\hux_{k\vert{k-1}}))\,\rg{\blam}_{t_k^\ttz}\,,
		\end{aligned}
	\end{equation}
	with the first term being the Jacobian of the sensing function $\ucb$ \wrt the kinematic quantity $\rg{\uscs}$ interpolated at $t_k^\ttz$ using the prior estimate $\hux_{k\vert{k-1}}$. The second term in \eqref{eq:hmat} refers to the Jacobian of the kinematic interpolation function $\rg{\uscs}$ w.r.t. the state vector. According to the linear expression given in \eqref{eq:kinoVec}, it follows
	\begin{equation*}
		\fJ_{\rg{\uscs}}(t_k^\ttz;\hux_{k\vert{k-1}})=\rg{\blam}_{t_k^\ttz}\,,
	\end{equation*}
	which is constant given the measurement timestamp $t_k^\ttz$ and a specified kinematic motion observed by the sensor (\algref{alg:filter}, line 9). Based on this, the prior estimate can be corrected by incorporating the measurement $\uz_k$ through a standard EKF update step, yielding the posterior estimate mean $\hux_{k\vert{k}}$ and covariance $\fP_{k\vert{k}}$ (\algref{alg:filter}, line 10--12).

	\subsection{Probabilistic Interpolation}\label{subsec:probItp}
	Embedding B-splines in recursive estimation decouples discrete-time state propagation from sensing continuous-time dynamical motions. Given the state estimate of mean $\hux_{k}$ and covariance $\fP_{k}$, we now aim to retrieve the motion estimates $\{\rg{\us}_{t_{i}}\}_{i=1}^m$ at arbitrary timestamps $\{t_i\}_{i=1}^m\subset(\,t_{n_{k-1}},t_{n_{k}}]$. For that, the queried motions are stacked into a vector
	\begin{equation}\label{eq:svec}
		\rg{\us}_{t_{1:m}}=[\,\rg{\us}_{t_1}^\top,\cdots,\rg{\us}_{t_{m}}^\top]^\top\in\R^{dm}\,,
	\end{equation}
	to which we perform \eqref{eq:kinoVec} in a batchwise manner. The kinematic interpolations are linear \wrt recurrent control points. Thus,  the mean and covariance of the motion estimates in \eqref{eq:svec} follow
	\begin{equation}\label{eq:probItp}
		\rg{\umu}_{t_{1:m}}=\rg{\blam}_{t_{1:m}}\hux_k\eqand\bSigma_{t_{1:m}}=\rg{\blam}_{t_{1:m}}\fP_{k}\,\rg{\blam}_{t_{1:m}}^\top\,,
	\end{equation}
	respectively, where the matrix
	\begin{equation*}
		\rg{\blam}_{t_{1:m}}=[\,\rg{\blam}_{t_1}^\top,\cdots,\rg{\blam}_{t_m}^\top]^\top\in\R^{dm\times{4d}}
	\end{equation*}
	denotes the combined coefficient matrices at each timestamp. Thus, we essentially obtain the joint probability distribution of the dynamical motion estimates at any timestamp, with different orders of temporal derivatives unified within a single computational procedure. To showcase the utility of the proposed SERE scheme, we now present a case study on motion tracking as follows.
	\begin{figure}[t!]
		\centering
		\adjustbox{trim={0.08\width} {0.1\height} {0.09\width} {0.01\height},clip}{\includegraphics[width=0.57\textwidth]{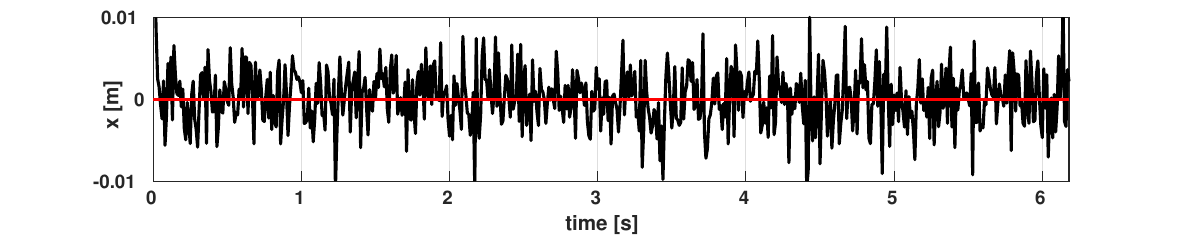}}\\
		\adjustbox{trim={0.08\width} {0.1\height} {0.09\width} {0\height},clip}{\includegraphics[width=0.57\textwidth]{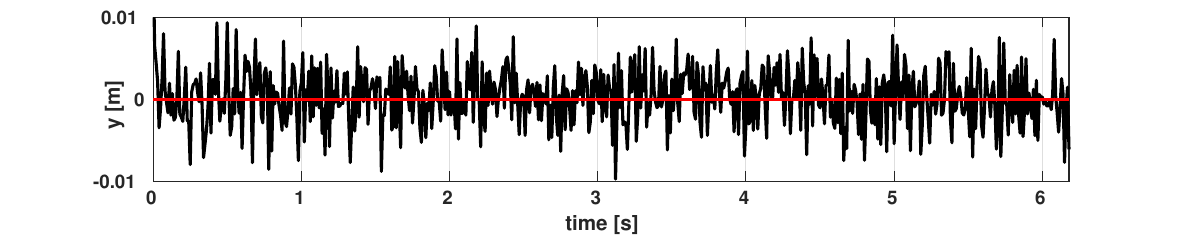}}\\
		\adjustbox{trim={0.08\width} {0\height} {0.09\width} {0.01\height},clip}{\includegraphics[width=0.567\textwidth]{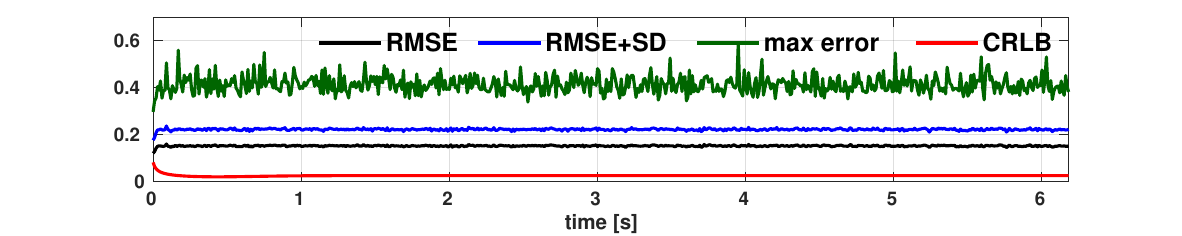}}\\
		\caption{Top and Middle: Averaged estimation errors (black) over the trajectory of \figref{fig:lissa}-(A) in $x$ and $y$ coordinates, respectively. Bottom: Overall RMSE compared to the CRLB, with the max error and RMSE plus one standard deviation (SD).}
		\label{fig:crlb}
	\end{figure}
	
	\subsection{Case Study}\label{subsec:case}
	We simulate a GPS-based tracking scenario along two-dimensional Lissajous curves with varying parameters~\cite{quad2021kanishke}, depicted as the red trajectories in \figref{fig:lissa}. The true trajectory is generated according to a constant-velocity model driven by noisy acceleration input as follows~\cite{gustafsson2010statistical} 
	\begin{equation*}
		\bbmat\,\us_{k+1}\\\dot{\us}_{k+1}\ebmat=\bbmat\fI_2 &T\fI_2 \\\fzero_2 &\fI_2\ebmat\bbmat\,\us_k\\\dot{\us}_k\ebmat+\bbmat\frac{T^2}{2}\fI_2\\T\fI_2\ebmat(\ua_k+\uw_k^\acc)\in\R^4\,,
	\end{equation*}
	where the process noise $\uw_k^\acc$ is simulated to be Gaussian-distributed with $\uw_k^\acc\sim\mN(\uzero_2,0.25\,\fI_2)$. The position $\us_k$ is observed via GPS at a frequency of \SI{100}{\hertz} (thus $T=0.01$\SI{}{\second}) under an additive Gaussian noise $\mN(\uzero_2,0.01\,\fI_2)$.
	
	We configure the proposed TriS model with a state vector $\ux_k\in\R^8$ concatenating RCPs $\{\uc_{n_k-i}\}_{i=0}^3\subset\R^2$ of knot interval $\tau$. The process model in \eqref{eq:sysCV} is employed for state propagation. The measurement model only exploits GPS sensing, and \eqref{eq:meas} is instantiated as 
	\begin{equation*}
		\uz_k={\blam}_{t_k^\ttz}\ux_k+\uv_k\,,\eqwith\blam_{t_k^\ttz}=\big(\bmo\,{\uu}_{t_k^\ttz}\big)^\top\otimes\fI_2
	\end{equation*}
	being the coefficient matrix for position interpolation at $t_k^\ttz\in(\,t_{n_{k-1}},t_{n_k}]$. $u_{t_k^\ttz}=(t_k^\ttz-t_{n_k-1})/\tau$ denotes the normalized timestamp according to \eqref{eq:u}. The measurement noise $\uv_k$ is set to be zero-mean with covariance $\fR_k=0.01\fI_2$.  Given the linear formulation above, the observation matrix in \eqref{eq:hmat} follows $\fH_k=\blam_{t_k^\ttz}$. We now provide several insights in the following discussions.
	
	\subsubsection{Qualitative validation}
	Across all four sequences, we set knot interval to $\tau=0.1$\SI{}{\second} and tune the zero-mean process noise in \eqref{eq:sysCV} with the covariance 
	$\fQ=(0.02\,\fI_6)\oplus(0.1\fI_2)$, where $\oplus$ denotes the direct sum as summarized in \secref{sec:notation}. As shown by the blue curves in \figref{fig:lissa}, the proposed SERE scheme delivers accurate tracking results along different trajectories. \figref{fig:lissasig} further illustrates the measurement uncertainty estimate along the trajectory in $x$ and $y$ coordinates on sequence \figref{fig:lissa}-(D) via the probabilistic interpolation according to \eqref{eq:probItp}. These qualitative results are evident to show the efficacy of the proposed SERE scheme, including the continuous-time probabilistic interpretation of the estimates.
	\begin{figure}[t!]
		\centering
		\adjustbox{trim={0.08\width} {0.1\height} {0.09\width} {0.01\height},clip}{\includegraphics[width=0.57\textwidth]{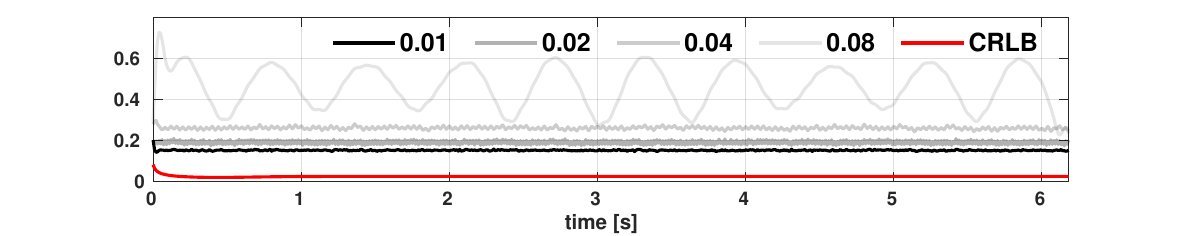}}\\
		\adjustbox{trim={0.08\width} {0\height} {0.09\width} {0\height},clip}{\includegraphics[width=0.57\textwidth]{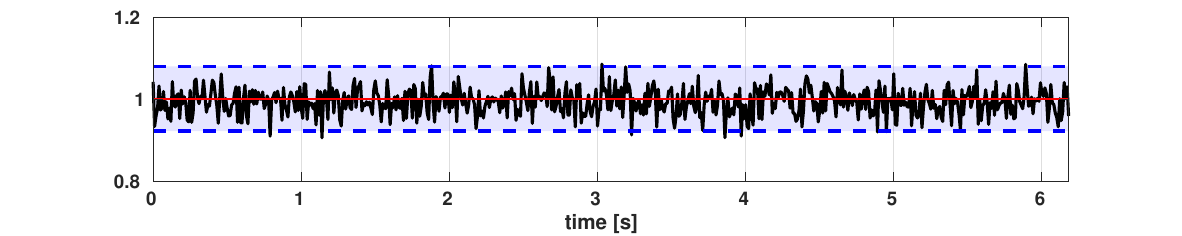}}\\
		\caption{Top: RMSEs for varying knot intervals $\tau$ compared with the CRLB on sequence \figref{fig:lissa}-(D). Bottom: ANEES result for knot interval $\tau=0.01$\SI{}{\second}. The blue shaded area represents the \SI{95}{\percent} confidence bounds.}
		\label{fig:anees}
	\end{figure}
	
	\begin{figure*}[t!]
		\centering
		\begin{tabular}{ccccc}
			&{{\textbf{\ttt{const1}}}} &{\textbf{\ttt{const2}}} &{\textbf{\ttt{const3}}} &{\textbf{\ttt{const4}}}\\
			\toprule
			\multirow[t]{1}{*}{\hspace{-1mm}\rotatebox{90}{~{\textbf{\ttt{tdoa2}}}}\hspace{-6mm}}&
			\adjustbox{trim={0.0\width} {0.0\height} {0.09\width} {0.0\height},clip}{\includegraphics[width=0.27\textwidth]{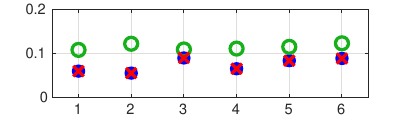}}&
			\adjustbox{trim={0.0\width} {0.0\height} {0.09\width} {0.0\height},clip}{\includegraphics[width=0.27\textwidth]{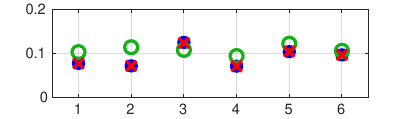}}&
			\adjustbox{trim={0.0\width} {0.0\height} {0.09\width} {0.0\height},clip}{\includegraphics[width=0.27\textwidth]{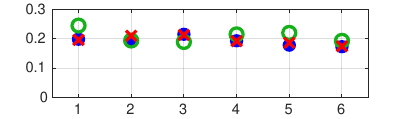}}&
			\adjustbox{trim={0.0\width} {0.0\height} {0.09\width} {0.0\height},clip}{\includegraphics[width=0.13\textwidth]{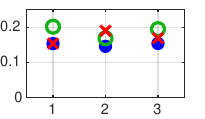}}\\
			\midrule
			\multirow[t]{1}{*}{\hspace{-1mm}\rotatebox{90}{~{\textbf{\ttt{tdoa3}}}}\hspace{-6mm}}&
			\adjustbox{trim={0.0\width} {0.0\height} {0.09\width} {0.0\height},clip}{\includegraphics[width=0.27\textwidth]{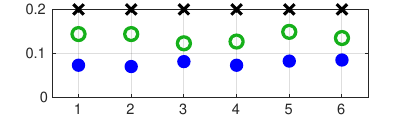}}&
			\adjustbox{trim={0.0\width} {0.0\height} {0.09\width} {0.0\height},clip}{\includegraphics[width=0.27\textwidth]{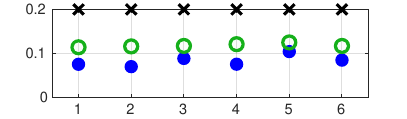}}&
			\adjustbox{trim={0.0\width} {0.0\height} {0.09\width} {0.0\height},clip}{\includegraphics[width=0.27\textwidth]{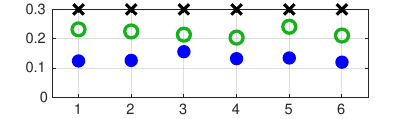}}&
			\adjustbox{trim={0.0\width} {0.0\height} {0.09\width} {0.0\height},clip}{\includegraphics[width=0.13\textwidth]{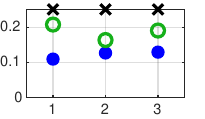}}\\
			\bottomrule
		\end{tabular}
		\caption{RMSEs in meters of TDoA-based motion tracking on the real-world dataset \ttt{UTIL} in Sec.\ref{sec:eva}-\ref{subsec:realEva}. Numbers on the horizontal axes denote \ttt{trial\#} in \ttt{const1} to \ttt{const3} and \ttt{trial1-traj\#} in \ttt{const4}. Results from SERE with outlier rejection are marked by {\color{blue}$\bullet$}. Results from SERE without outlier rejection and from the baseline method are denoted by {\color{Red}$\pmb{\pmb\times}$} and {\color{mGreen}{$\pmb{\pmb{\circ}}$}}, respectively. Without outlier rejection, SERE frequently exhibits tracking failures under the distributed TDoA mode, as indicated by {\color{Black}$\pmb{\pmb\times}$} in \ttt{tdoa3}.}
		\label{fig:tdoaEva}
	\end{figure*}
	
	\subsubsection{Statistical investigation}
	We further configure the proposed SERE scheme with a process noise covariance of $\fQ=(0.5\,\fI_6)\oplus\fI_2$ and evaluate its performance in terms of statistical efficiency and consistency through $1000$ Monte Carlo (MC) runs. The trajectory shown in \figref{fig:lissa}-(A) is first selected, where we run SERE with knot interval $\tau=\,$\SI{0.01}{\second}. As illustrated in \figref{fig:crlb} through the averaged error in the $x$ and $y$ coordinates, the configured SERE scheme clearly demonstrates unbiased estimation performance. We also compute the overall root mean squared error (RMSE) across MC runs and compare it with the Cramér–Rao lower bound (CRLB)~\cite{bergman1997point,gustafsson2010statistical}. Under the current configuration, our estimator exhibits limited statistical efficiency, as indicated by the offset between RMSE and the CRLB. 
	
	To further investigate the impact of knot interval $\tau$ on tracking performance, we select sequence \figref{fig:lissa}-(D), which features more dynamic motion patterns. As shown in the top plot of \figref{fig:anees}, reducing the knot interval estimation can improve both the accuracy and statistical efficiency of the estimator, up to a certain extent. Additionally, we extract the result for $\tau=0.01$ and compute the average normalized estimation error squared (ANEES) to assess the consistency (credibility) of the estimator~\cite{li2001practical}
	\begin{equation*}
		\epsilon=\frac{1}{dm}\sum_{i=1}^{m}(\us_i-\hus_i)\bSigma_i^{-1}(\us_i-\hus_i)\,, 
	\end{equation*}
	where $\hus_i$ and $\bSigma_i$ are the interpolated position and covariance estimates according to \eqref{eq:probItp}. $d=2$ and $m=1000$ denote the dimension and number of MC runs, respectively. As shown in the bottom of \figref{fig:anees}, the configured estimator demonstrates consistency, as indicated by its ANEES  remaining around $1$ within \SI{95}{\percent} confidence interval.
	
	\subsubsection{Brief summary} The presented case study validates the effectiveness of the proposed SERE scheme for motion tracking and its underlying probabilistic soundness. When properly configured, SERE reliably produces unbiased and consistent estimates. Due to the approximative nature of B-splines in modeling system dynamics, SERE may exhibit a limitation in statistical efficiency relative to the theoretical optimum established by standard discrete-time state-space modeling. Moreover, tuning the parameters, particularly, the knot interval and the process noise for RCPs propagation, plays crucial rules in estimation performance. These aspects will be further discussed in Sec.\ref{sec:eva}-\ref{subsec:discuss}, in the context of multisensor motion tracking.
	\begin{figure*}[t!]
		\centering
		\begin{tabular}{cccc}
			\adjustbox{trim={0.03\width} {0.07\height} {0.03\width} {0.14\height},clip}{\includegraphics[width=0.243\textwidth]{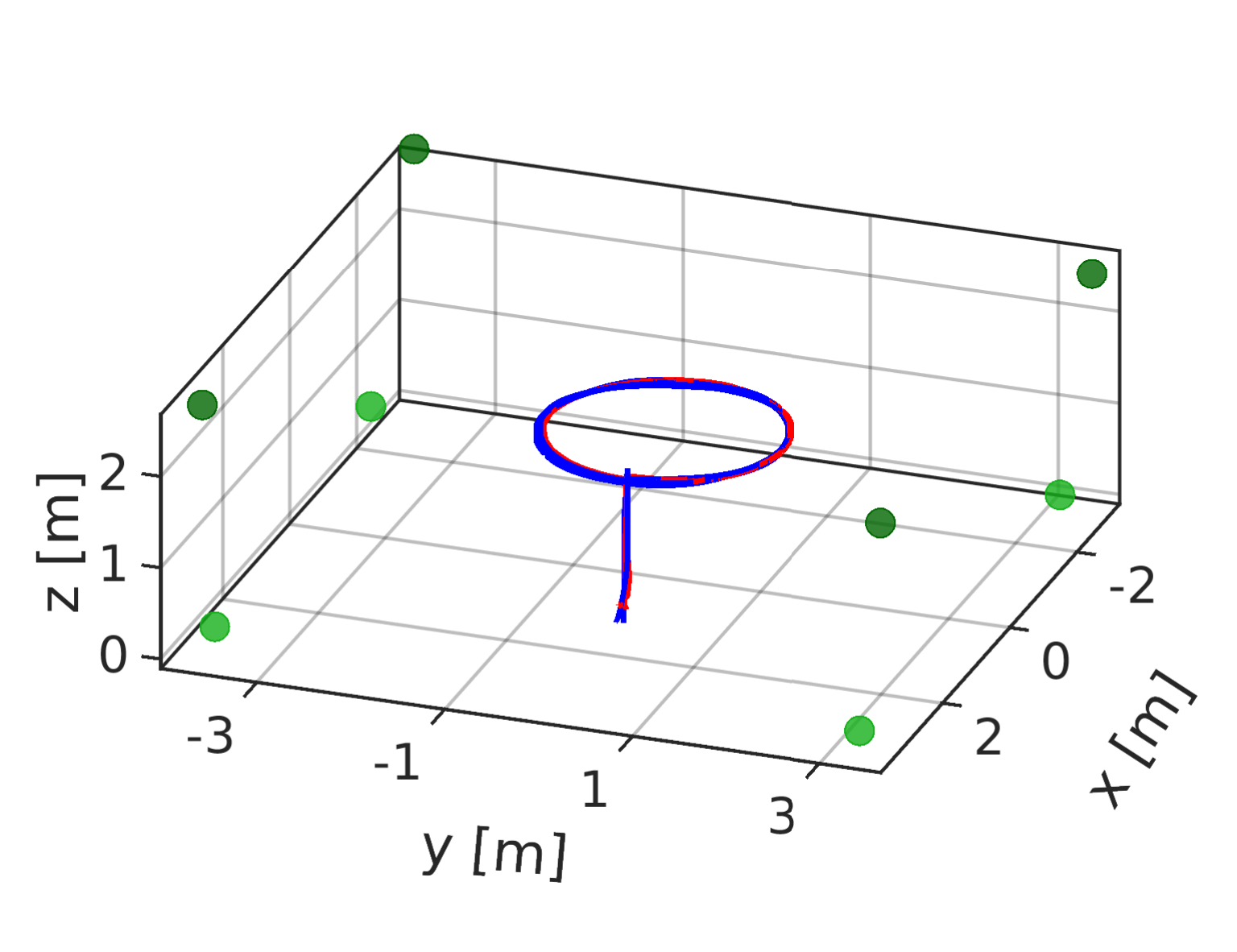}} &
			\adjustbox{trim={0.03\width} {0.07\height} {0.03\width} {0.14\height},clip}{\includegraphics[width=0.243\textwidth]{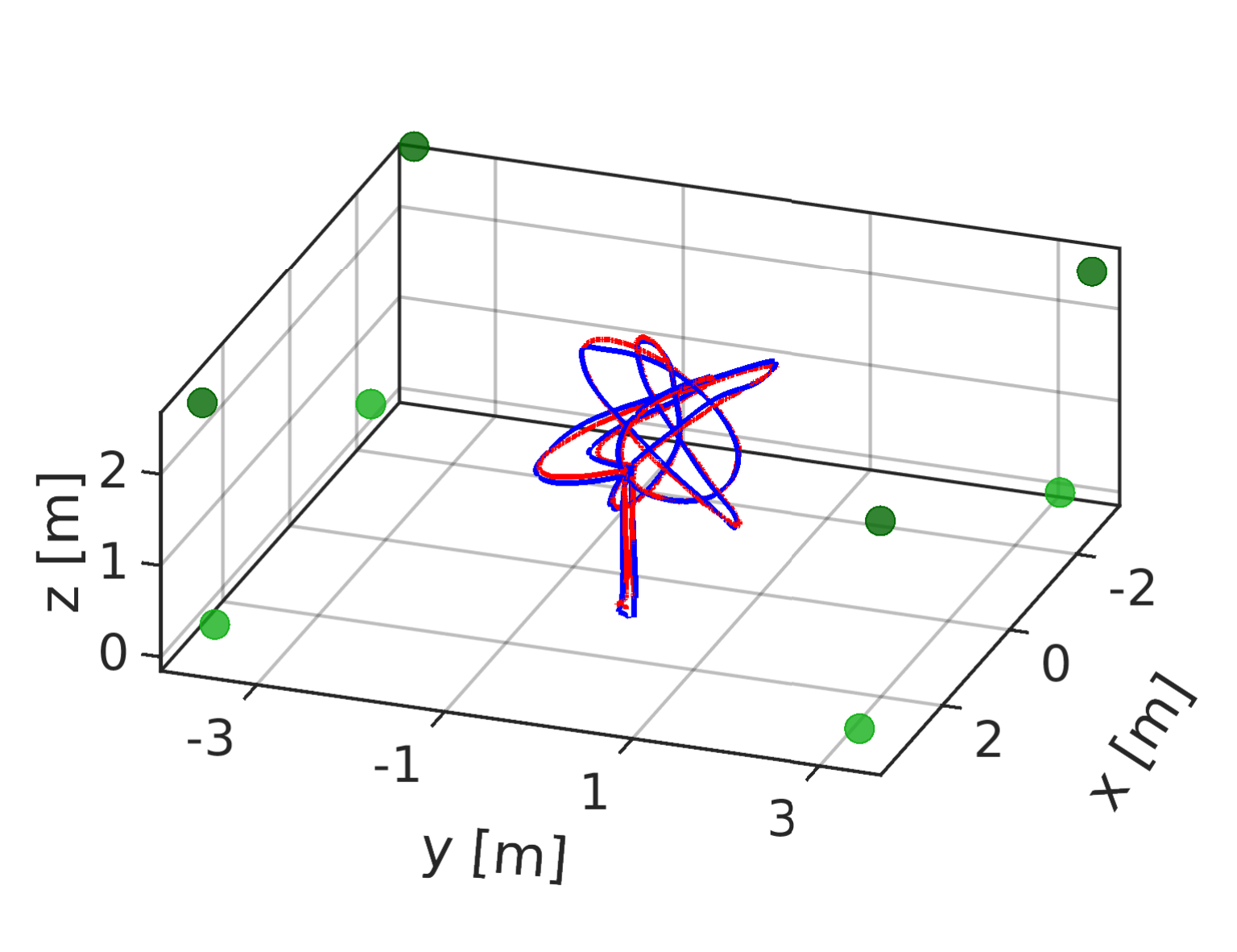}} &
			\adjustbox{trim={0.03\width} {0.07\height} {0.03\width} {0.14\height},clip}{\includegraphics[width=0.243\textwidth]{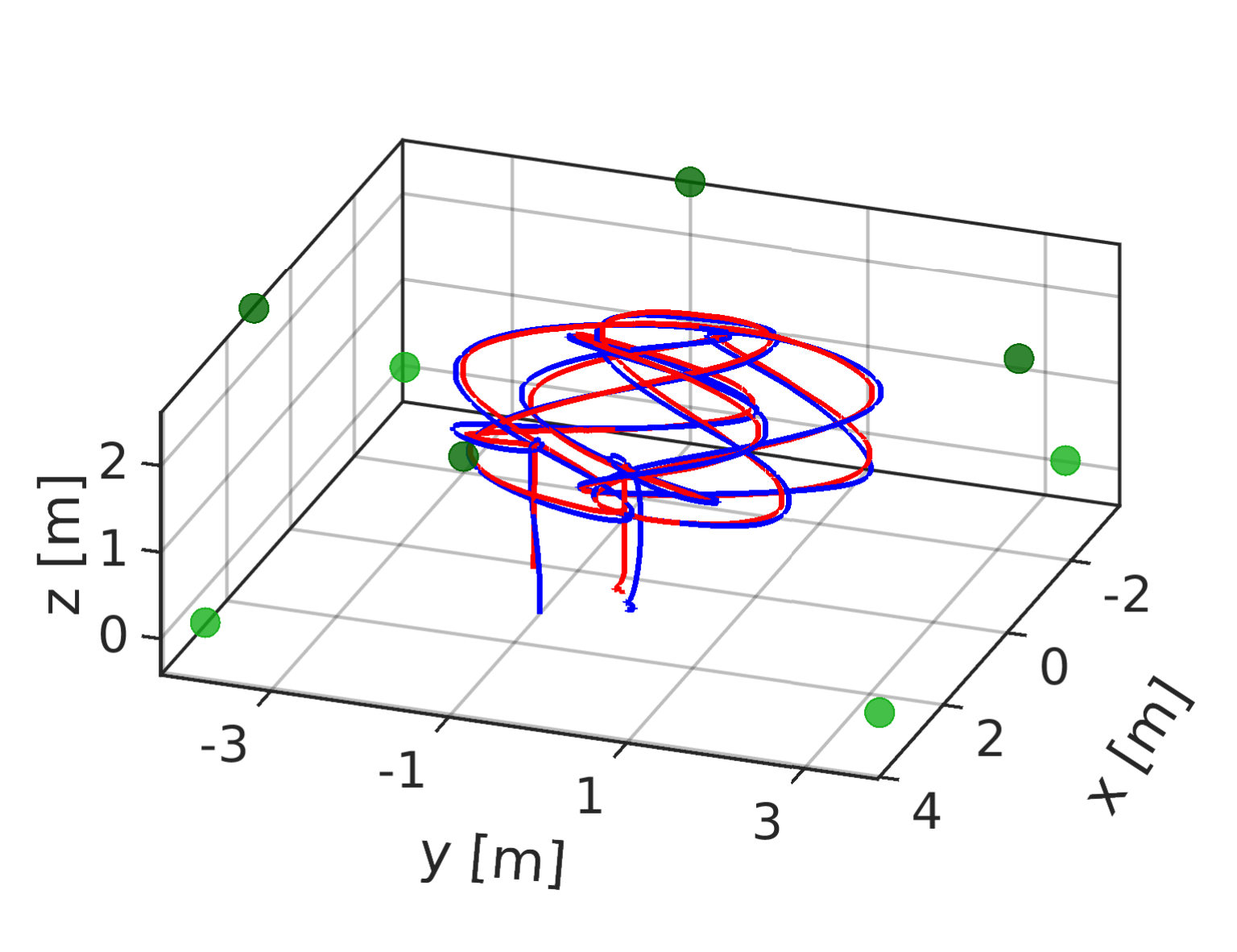}} &
			\adjustbox{trim={0.03\width} {0.07\height} {0.03\width} {0.14\height},clip}{\includegraphics[width=0.243\textwidth]{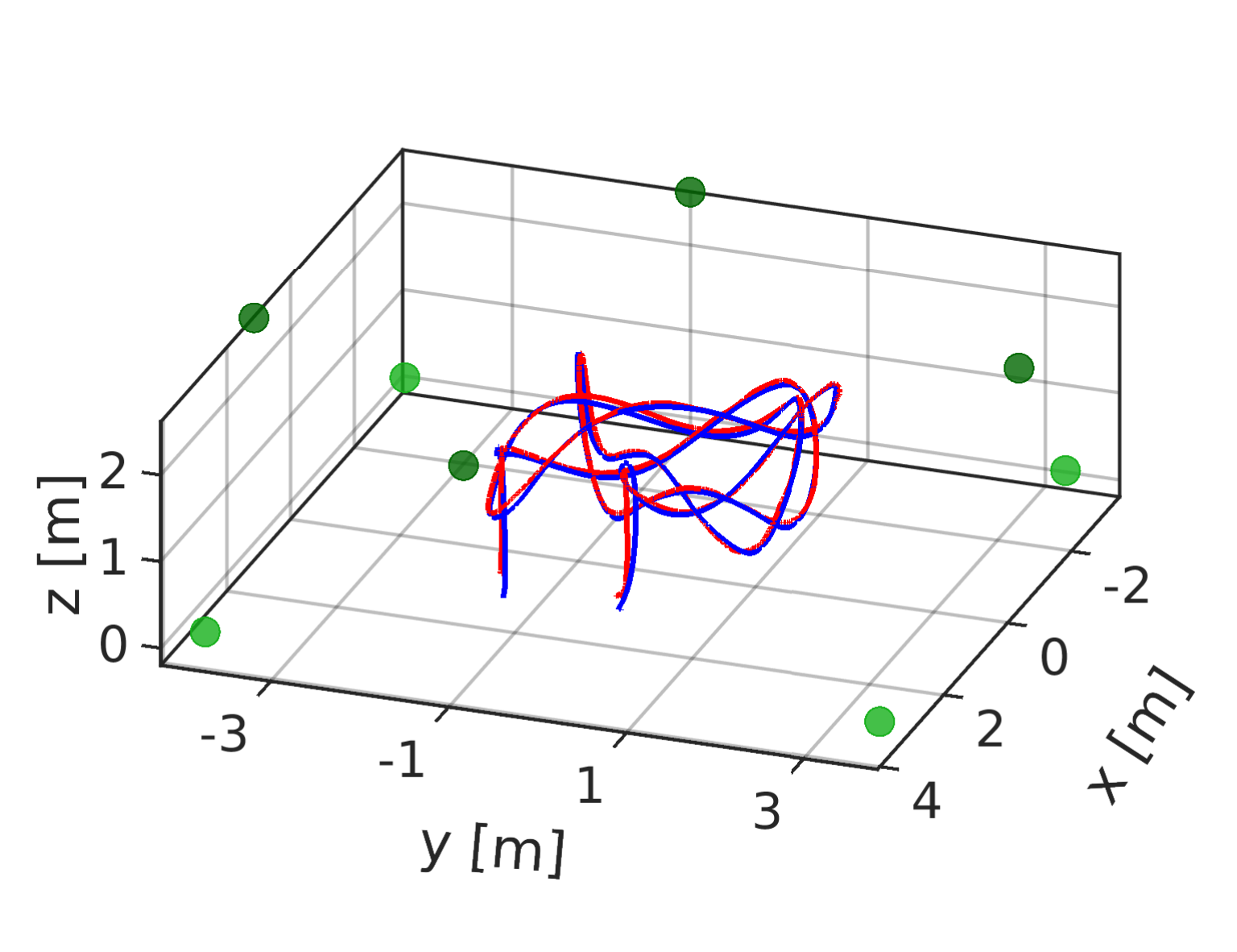}} \\
			\subcap{\ttt{tdoa2-const1-trial1}} &\subcap{\ttt{tdoa2-const1-trial2}} &\subcap{\ttt{tdoa2-const2-trial3}} &\subcap{\ttt{tdoa2-const2-trial4}} \\
			\\
			\adjustbox{trim={0.03\width} {0.04\height} {0.03\width} {0.14\height},clip}{\includegraphics[width=0.243\textwidth]{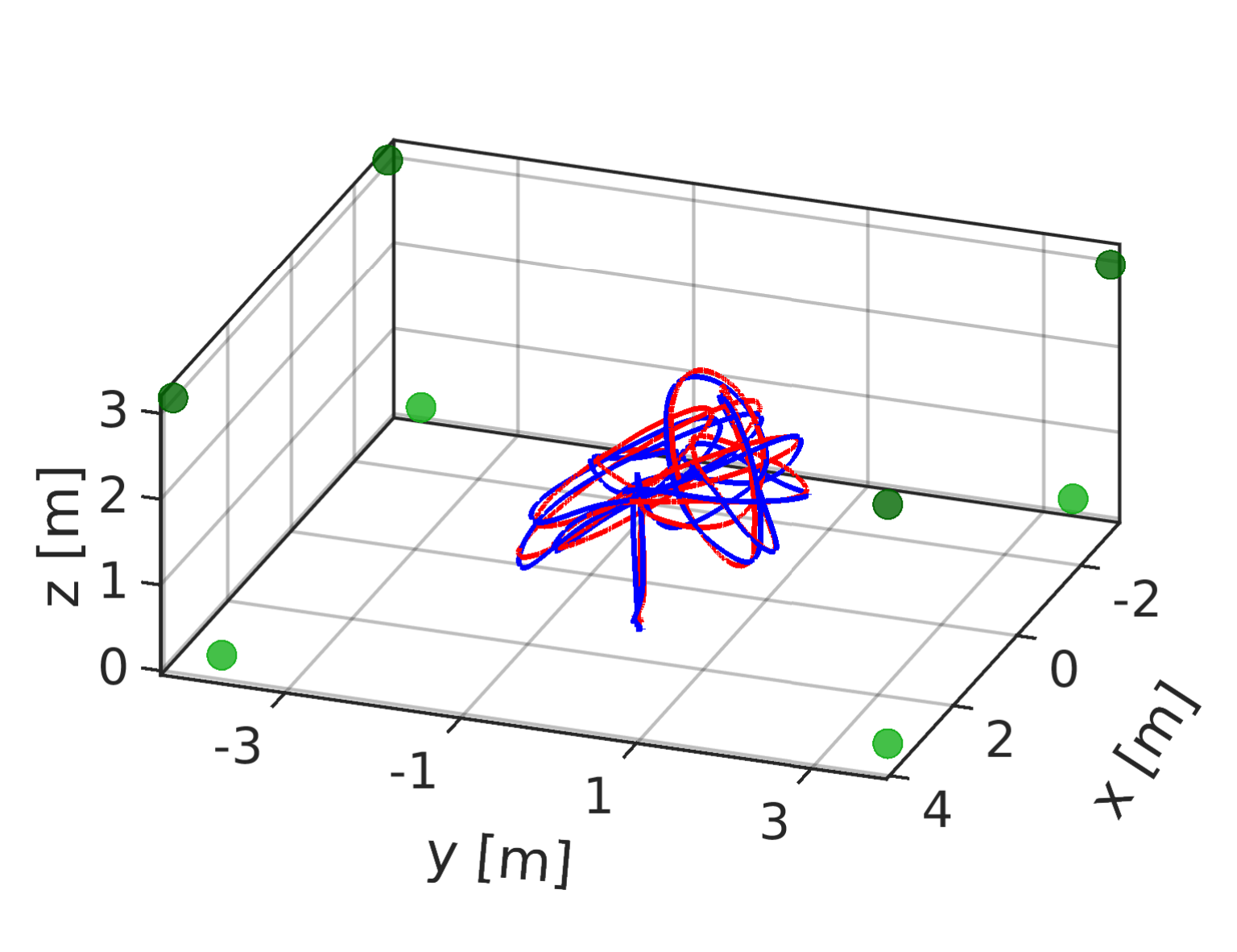}} &
			\adjustbox{trim={0.03\width} {0.04\height} {0.03\width} {0.14\height},clip}{\includegraphics[width=0.243\textwidth]{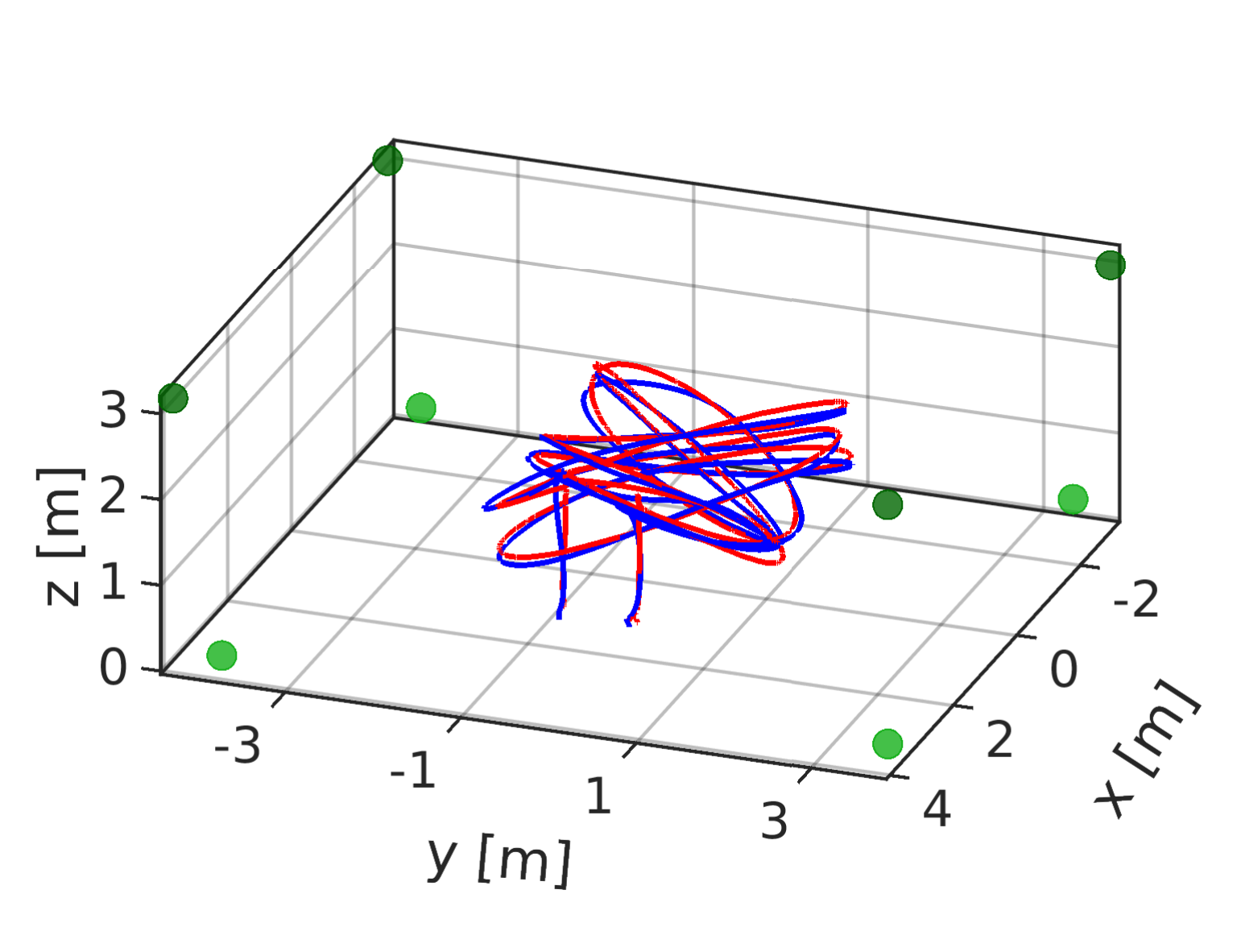}} &
			\adjustbox{trim={0.03\width} {0.04\height} {0.03\width} {0.14\height},clip}{\includegraphics[width=0.243\textwidth]{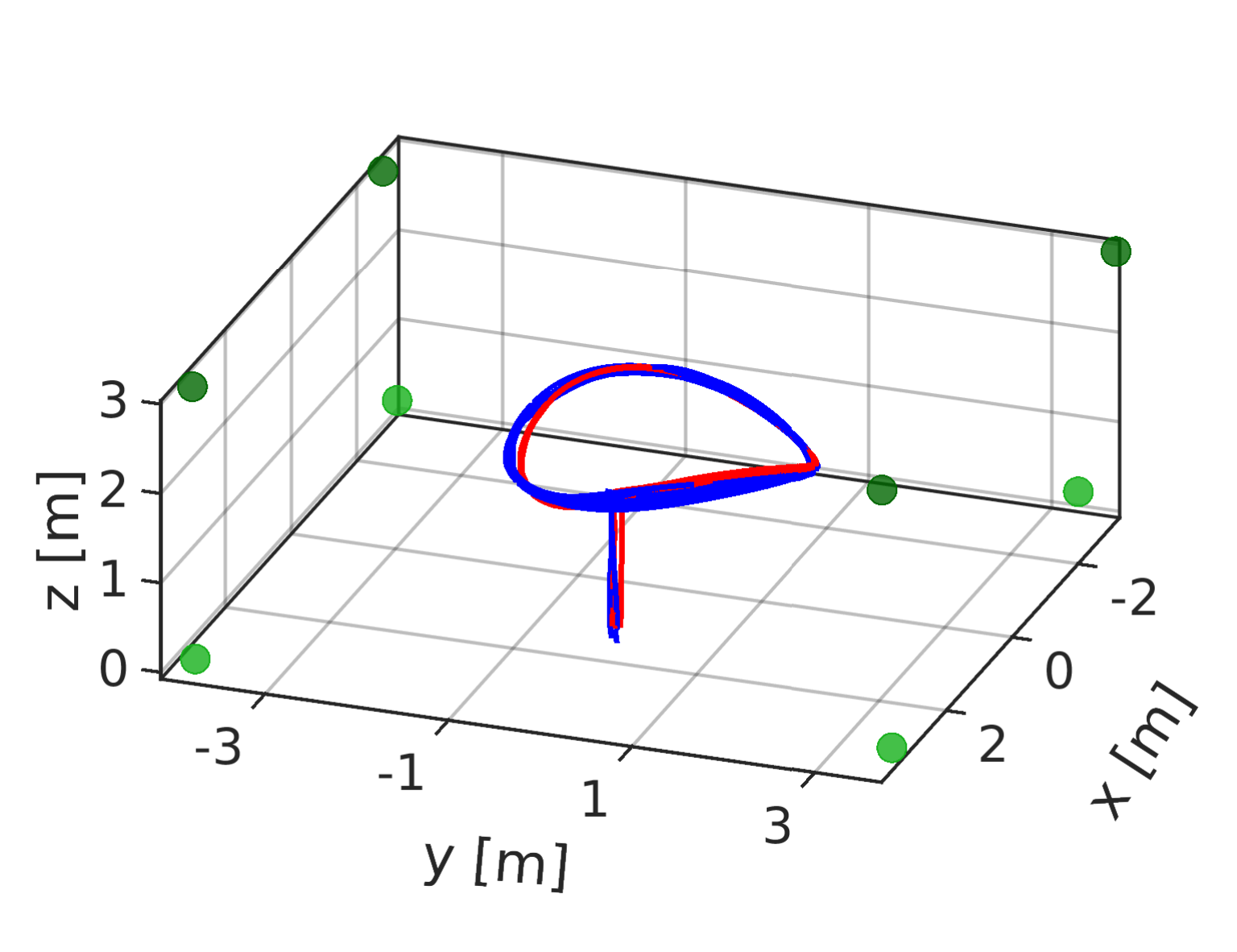}} &
			\adjustbox{trim={0.03\width} {0.04\height} {0.03\width} {0.14\height},clip}{\includegraphics[width=0.243\textwidth]{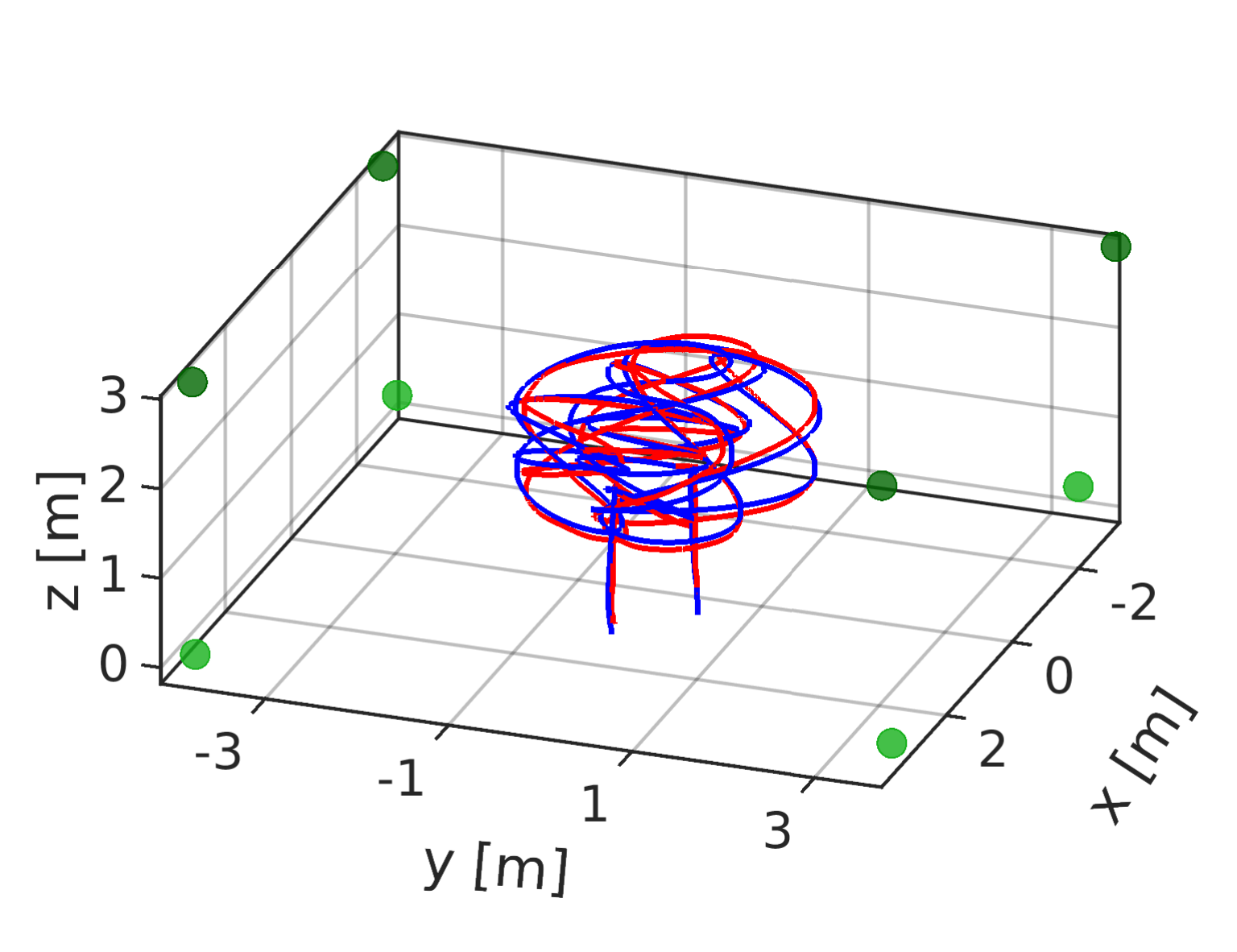}} \\
			\subcap{\ttt{tdoa3-const3-trial5}} &\subcap{\ttt{tdoa3-const3-trial6}} &\subcap{\ttt{tdoa3-const4-traj2$^\ttt{*}$}} &\subcap{\ttt{tdoa3-const4-traj3$^\ttt{*}$}}\\
		\end{tabular}
		\caption{Representative TDoA-only tracking results given by SERE on \ttt{UTIL} in Sec.\ref{sec:eva}-\ref{subsec:realEva}, illustrating all shapes of trajectories, TDoA modes, and anchor constellations. Blue and red curves depict the estimates and ground truth, respectively, and green dots denotes UWB anchors. ($\ttt{*}$: sequences of \ttt{trial1})}
		\label{fig:tdoaPic}
	\end{figure*}
	
	\section{\uppercase{Evaluation}}\label{sec:eva}
	We now conduct an in-depth evaluation of the proposed SERE scheme, covering both time-difference-of-arrival (TDoA) and time-of-arrival (ToA)-inertial settings based on real-world scenarios\footnote{Both TDoA and ToA measurements are expressed as distance-equivalent quantities.}. Comparisons with conventional recursive filtering methods are included.
	
	\subsection{Tracking Scenarios and Settings}\label{subsec:trackEva}
	We exploit the \ttt{UTIL} dataset for UWB-based 3D position tracking in sensor networks~\cite{zhao2022utilee}. The original \ttt{UTIL} dataset is recorded onboard a quadrotor platform equipped with a UWB tag and an IMU as it flies along various trajectories~\cite{zhao2022utilee,RAL23_Li}. The low-cost UWB sensor network is configured using multiple anchor constellations (\ttt{const1-4}) and operates at both centralized (\ttt{tdoa2}) and decentralized (\ttt{tdoa3}) TDoA modes. Sensor measurements are received at varying rates, with the UWB rate ranging from \SI{200}{\hertz} to \SI{500}{\hertz} and the IMU at approximately \SI{1000}{\hertz}. In the remainder of this section, we base our evaluation on overall 42 sequences from the original dataset of diverse trajectories, covering both TDoA modes and all types of constellations. 
		
	\subsubsection{Real-world scenario of TDoA sensing}
	For the real-world evaluation in Sec.\ref{sec:eva}-\ref{subsec:realEva}, we adopt the original TDoA measurements from the selected sequences of \ttt{UTIL}. The TDoA data are affected by well-known issues in UWB ranging and tracking, such as non-line-of-sight and multi-path conditions~\cite{kok2015indoor}. Additionally, obstacles of various types of materials, including ferromagnetic ones, introduce further complexity by interfering with signal propagation. In the centralized mode \ttt{tdoa2}, anchors are globally synchronized, ensuring temporally ordered TDoA measurements overtime. However, in the decentralized mode \ttt{tdoa3}, anchors are randomly paired without global clock synchronization, leading to unknown time offset in TDoA measurements. These intertwined factors give rise to complex noise patterns and outliers, posing substantial challenges for TDoA-only motion tracking~\cite{RAL23_Li,kok2015indoor,zhao2022utilee}. 
		
	\subsubsection{Synthetic scenario of ToA-inertial sensing}\label{subsubsec:toa}
	Using the selected sequences from \ttt{UTIL}, we synthesize a ToA-inertial tracking scenario for the evaluation in Sec.\ref{sec:eva}-\ref{subsec:synEva}. The ground truth trajectories are used to simulate the accelerometer w.r.t.\ the world frame and ToA observations\footnote{We simulate the accelerometer in the global frame (instead of the IMU body frame as in reality) to bypass the need for orientation estimation, which lies outside the scope of this article.}. Both sensors are sampled at the original IMU and UWB timestamps provided in \ttt{UTIL}. Moreover, the ToA and inertial observations are further corrupted by zero-mean Gaussian noises with covariances  $\text{R}^\toa=0.01$ and $\fR^\acc=0.01\,\fI_3$, respectively. In the subsequent evaluation, we refer to this synthetic dataset as \ttt{Syn-UTIL}. 
	
	\subsection{Real-World Evaluation on TDoA Motion Tracking}\label{subsec:realEva}
	We configure the proposed SERE scheme using control points of knot interval $\tau=\,$\SI{2}{\second}. The state vector $\ux_k\in\R^{12}$ composing RCPs $\{\uc_{n_k-i}\}_{i=0}^3\subset\R^3$ evolves according to the process model in \eqref{eq:sysCV} with noise covariance $\fQ=0.01\fI_{12}$. The following measurement model is set up according to TDoA ranging
	\begin{equation}\label{eq:tdoa}
		z_k^\tdoa=\Vert\blam_{t_k^\ttz}\ux_k-\uio_k^i\Vert-\Vert\blam_{t_k^\ttz}\ux_k-\uio_k^j\Vert+v_k^\tdoa\,,
	\end{equation}
	where $\uio_k^i,\uio_k^j\in\R^3$ denote the coordinates of the two connected anchors. $v_k^\tdoa$ is the zero-mean measurement noise with variance $\fR_k=0.05$. Accordingly, the observation matrix in \eqref{eq:hmat} for update is specified as 
	\begin{equation*}
		\fH_k=\bigg(\frac{(\blam_{t_k^\ttz}\hux_{k\vert{k-1}}-\uio_{k}^i)^\top}{\Vert\blam_{t_k^\ttz}\hux_{k\vert{k-1}}-\uio_{k}^i\Vert}-\frac{(\blam_{t_k^\ttz}\hux_{k\vert{k-1}}-\uio_{k}^j)^\top}{\Vert\blam_{t_k^\ttz}\hux_{k\vert{k-1}}-\uio_{k}^j\Vert}\bigg)\blam_{t_k^\ttz}\,.
	\end{equation*}
	
	\textit{Outlier rejection:\,} During the update step, we compute the Mahalanobis distance of TDoA measurement $z_k^\tdoa$ w.r.t.\ the predicted value, and accept the measurement if the distance falls below a predefined threshold, namely,
	\begin{equation}\label{eq:rej}
		\frac{(z_k^\tdoa-\sch_k(\hux_{k\vert{k-1}}))^2}{\fH_k\fP_{k\vert{k-1}}\fH_k^\top+\fR_k}<\epsilon^2_\tdoa\,.
	\end{equation}
	where $\hux_{k\vert{k-1}}$ denotes the predicted prior, and the threshold is set to $\epsilon^2_\tdoa=15$. The denominator term can be directly obtained via the probabilistic interpolation \eqref{eq:probItp}. 
	
	To quantify the TDoA-only tracking accuracy of SERE, we compute the RMSE of the absolute position error (APE) over the entire trajectory, where estimates are interpolated at the ground truth timestamps (\SI{200}{\hertz}) via \eqref{eq:kino0}. To showcase SERE's probabilistic capability in handling real-world data, we evaluate the scheme both with and without the outlier rejection mechanism in \eqref{eq:rej}. For comparison with a conventional discrete-time estimator, we include the baseline approach provided by the \ttt{UTIL} dataset, namely, a $6$-DoF error-state Kalman filter (ESKF) using both TDoA and inertial measurements (including outlier rejection based on predicted states through inertial propagation)~\cite{zhao2022utilee}.
	
	As shown in \figref{fig:tdoaEva}, the proposed SERE scheme with outlier rejection delivers consistently superior results in terms of tracking accuracy and robustness over ESKF using TDoA-inertial setting. The proposed spline-embedded outlier rejection mechanism can effectively improve the tracking robustness under challenging conditions, as evidenced by the results from sequences in \ttt{tdoa3}.
	
	\begin{figure*}[t]
		\centering
		\begin{tabular}{ccccc}
			&{\textbf{\ttt{const1}}} &\textbf{\ttt{const2}} &\textbf{\ttt{const3}} &\textbf{\ttt{const4}}\\
			\toprule
			\multirow[t]{1}{*}{\hspace{-1mm}\rotatebox{90}{~\textbf{\ttt{toa2}}}\hspace{-6mm}}
			&\adjustbox{trim={0.045\width} {0.035\height} {0.09\width} {0.0\height},clip}{\includegraphics[width=0.28\textwidth]{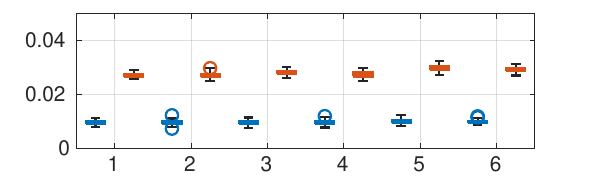}}
			&\adjustbox{trim={0.045\width} {0.035\height} {0.09\width} {0.0\height},clip}{\includegraphics[width=0.28\textwidth]{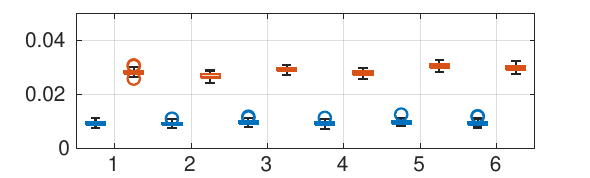}}
			&\adjustbox{trim={0.045\width} {0.035\height} {0.09\width} {0.0\height},clip}{\includegraphics[width=0.28\textwidth]{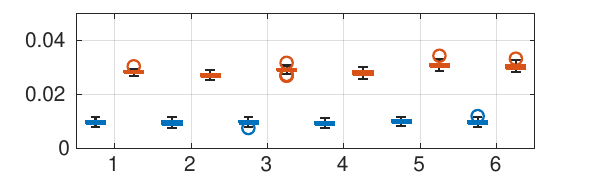}}
			&\adjustbox{trim={0.003\width} {0.035\height} {0.09\width} {0.0\height},clip}{\includegraphics[width=0.14\textwidth]{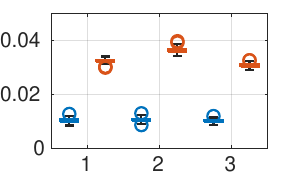}}\\
			\midrule
			\multirow[t]{1}{*}{\hspace{-1mm}\rotatebox{90}{~\textbf{\ttt{toa3}}}\hspace{-6mm}}
			&\adjustbox{trim={0.045\width} {0.035\height} {0.09\width} {0.0\height},clip}{\includegraphics[width=0.28\textwidth]{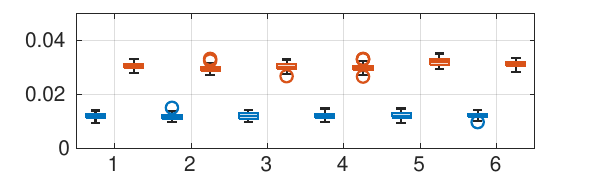}}
			&\adjustbox{trim={0.045\width} {0.035\height} {0.09\width} {0.0\height},clip}{\includegraphics[width=0.28\textwidth]{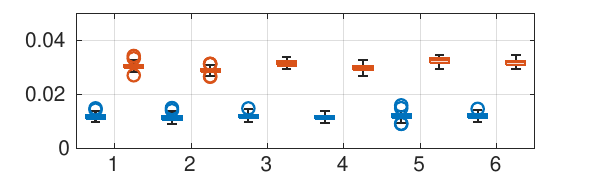}}
			&\adjustbox{trim={0.045\width} {0.035\height} {0.09\width} {0.0\height},clip}{\includegraphics[width=0.28\textwidth]{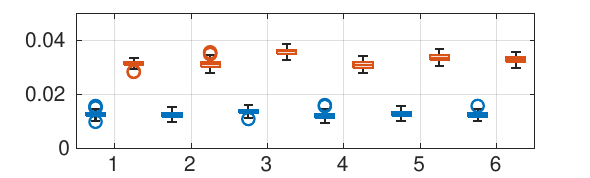}}
			&\adjustbox{trim={0.003\width} {0.035\height} {0.09\width} {0.0\height},clip}{\includegraphics[width=0.14\textwidth]{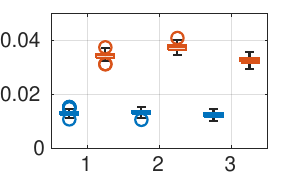}}\\
			\bottomrule
		\end{tabular}
		\caption{Results of ToA-inertial tracking on the synthetic dataset \ttt{Syn-UTIL} in Sec.\ref{sec:eva}-\ref{subsec:synEva}. The vertical axes denote RMSEs \wrt ground truth, and their distributions on each sequence are plotted using the \ttt{boxchart} function of default setting in \ttt{MATLAB}. SERE (blue) delivers consistently superior accuracy over the baseline method (red).}
		\label{fig:toaEva}
	\end{figure*}
	
	\subsection{Evaluation on ToA-Inertial Motion Tracking}\label{subsec:synEva}
	We configure the SERE scheme with a knot interval of $\tau=$\,\SI{1}{\second} and adopt the same type of process model in Sec.\ref{sec:eva}-\ref{subsec:realEva}. The process noise covariance is set to
	\begin{equation*}
		\fQ=(0.02\times10^{-5}\,\fI_{9})\oplus(10^{-2}\,\fI_3)\,.
	\end{equation*}
	Given the multimodal setting, the following measurement models are formulated according to \eqref{eq:meas}
	\begin{equation*}
		\uz_k^\acc=\ddot{\blam}_{t_k^\ttz}\ux_k+\uv_k^\acc\eqand z_k^\toa=\Vert\blam_{t_k^\ttz}\ux_k-\uio_k\Vert+v_k^\toa\,,
	\end{equation*}
	corresponding to an accelerometer or a ToA measurement at timestamp $t_k^\ttz$, respectively. $\uio_k$  denotes the indexed anchor coordinates. The noise terms $\uv_k^\acc$ and $v_k^\toa$ are zero-mean with covariances $\fR^\acc$ and $\fR^\toa$ as described in Sec.\ref{sec:eva}-\ref{subsec:trackEva}.\ref{subsubsec:toa}, respectively. Corresponding observation matrices following \eqref{eq:hmat} can be derived as follows
	\begin{equation*}
		\fH_k^\acc=\ddot{\blam}_{t_k^\ttz} \eqand
		\fH_k^\toa=\frac{(\blam_{t_k^\ttz}\hux_{k\vert{k-1}}-\uio_{k})^\top\blam_{t_k^\ttz}}{\Vert\blam_{t_k^\ttz}\hux_{k\vert{k-1}}-\uio_{k}\Vert}\,.
	\end{equation*} 
	
	Similar to the real-world evaluation in Sec.\ref{sec:eva}-\ref{subsec:realEva}, we compute the RMSE of position estimates \wrt the ground truth across the trajectory. For comparison to conventional recursive estimation scheme, we implement an EKF as the baseline with a state vector comprising the position and velocity. The prediction and update steps are scheduled asynchronously according to the multisensor fusion scenario. Similarly to the simulation setup in Sec.\ref{sec:filter}-\ref{subsec:case}, we propagate the state according to a constant-velocity model with the acceleration input. Afterward, an EKF update step is performed to fuse the range measurement and deliver the posterior estimate. Note that, in this synthetic scenario, no outlier rejection is necessary.
	
\begin{figure}[t]
		\centering
		\begin{tabular}{cc}
			\adjustbox{trim={0.01\width} {0.03\height} {0.08\width} {0.11\height},clip}{\includegraphics[width=0.22\textwidth]{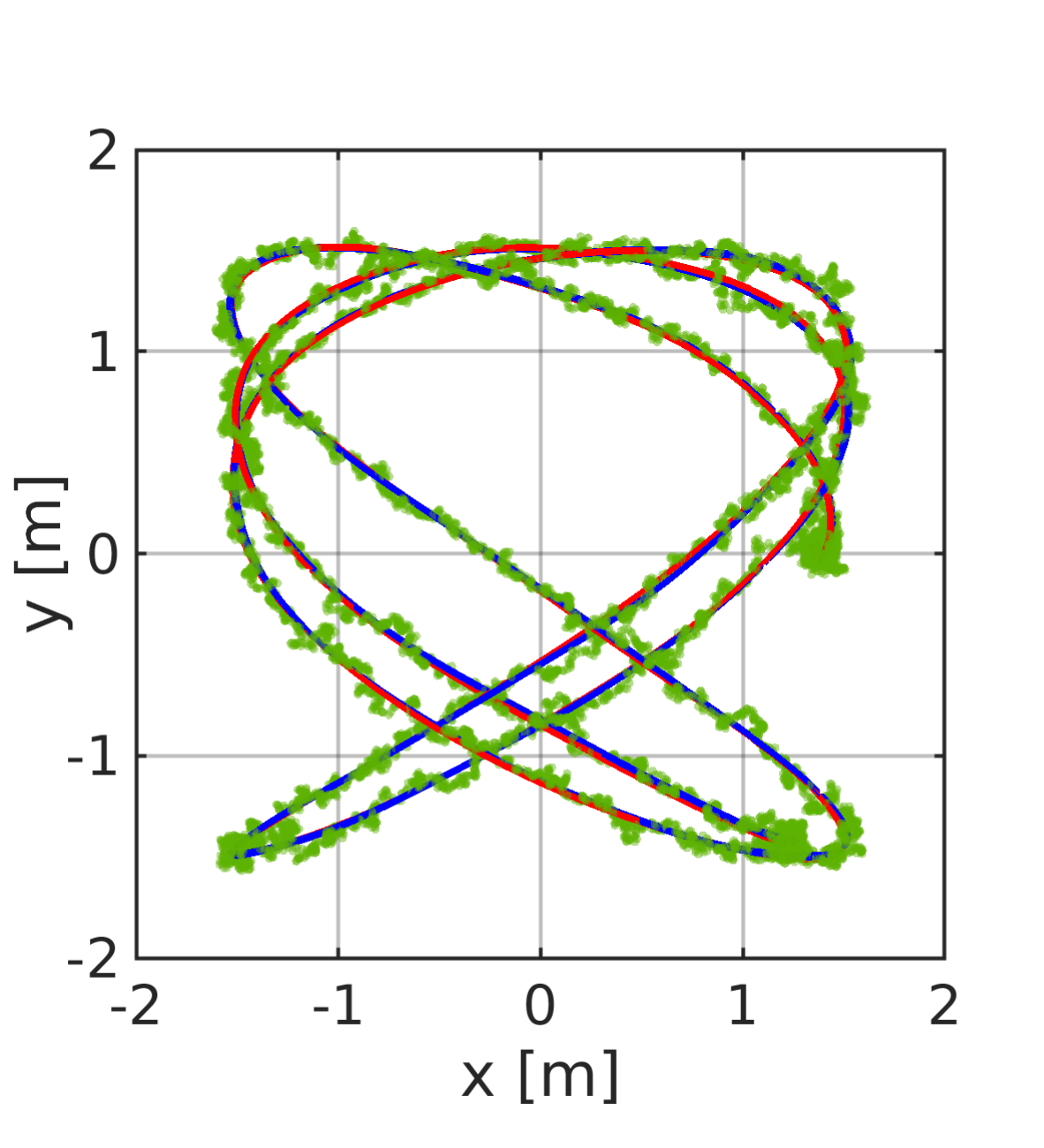}} &
			\adjustbox{trim={0.01\width} {0.03\height} {0.08\width} {0.11\height},clip}{\includegraphics[width=0.22\textwidth]{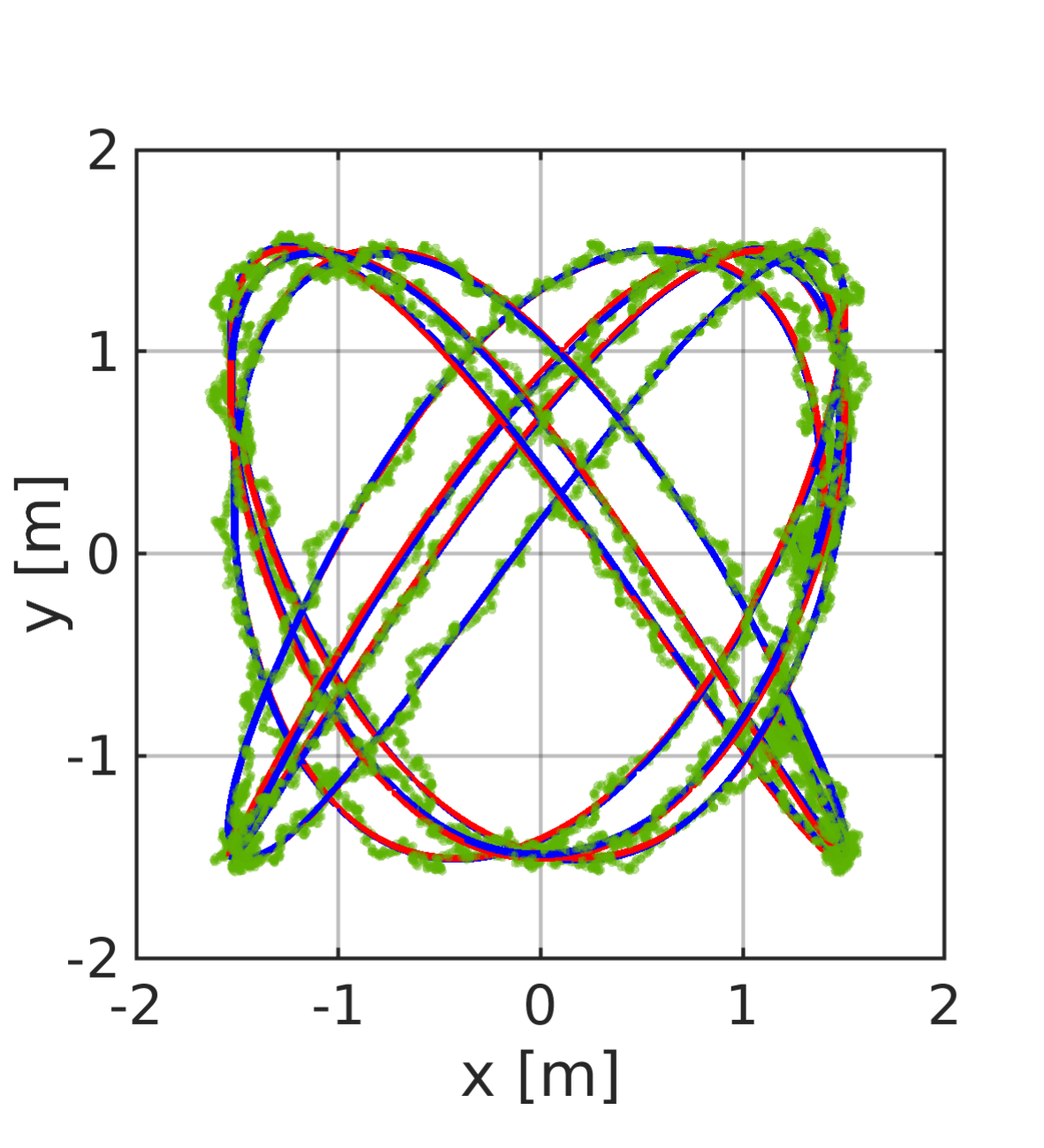}}\\
			\subcap{\ttt{const2-trial4}} & \subcap{\ttt{const2-trial6}}
		\end{tabular}
		\caption{Trajectory estimates from the proposed (blue) and conventional (green) schemes, shown against ground truth (red) on two sequences from \ttt{Syn-UTIL-tdoa2} with increased ToA measurement noise levels.} 
		\label{fig:comp}
	\end{figure}

	We perform overall $100$ Monte Carlo runs on each sequence of \ttt{Syn-UTIL}, and the resulting RMSE distributions are summarized in \figref{fig:toaEva}. The proposed spline embedding enables considerable improvement on tracking accuracy across all sequences compared with the discrete-time counterpart. For demonstration, a tracking animation on sequence \ttt{const2-trial3-tdoa3} is provided\footnote{Please visit: \href{https://www.youtube.com/watch?v=RJYDkYaBFs0}{https://www.youtube.com/watch?v=RJYDkYaBFs0}}. To further highlight the strength of spline embedding in noise adaptation, we perform additional tests on \ttt{trial4} and \ttt{trial6} of \ttt{tdoa2} in \ttt{Syn-UTIL} with increased noise in ToA ranging ($\text{R}^\toa=0.1$). As shown by the top-view plots in \figref{fig:comp}, the proposed SERE scheme produces accurate trajectory estimates with inherent kinematic consistency thanks to the B-spline embedding. In contrast, the baseline exhibits physically infeasible motion transitions due to the conventional state-space formulation. 
	
	\subsection{Discussion}\label{subsec:discuss}
	Achieving high performance in state estimation using spline embedding requires proper configurations of the system. In addition to the insights provided by the case study in Sec.\ref{sec:filter}-\ref{subsec:case}, we now discuss the tuning of two key parameters as follows.
		
	\textit{Temporal (knot) interval:\, } The knot interval $\tau$ of the control points is a global parameter in spline embedding and facilitates efficient state storage. In both evaluation scenarios, SERE only needs to store the control points at a frequency of \SI{1}{} or \SI{2}{\hertz}, while still delivering trajectory estimates as continuous functions over time. In contrast, conventional methods store discrete-time states at the rate of asynchronous measurements (between $200$ and $1000$ Hz). In principle, a smaller value of $\tau$ allows for more detailed modeling of dynamical motions. However, this also requires higher rate of sensor measurements for more frequent state propagation and correspondingly more memory consumption for state storage. 
	
	\textit{Process noise:\, } Tuning the process noise in deploying the SERE scheme follows common practice for conventional stochastic filtering. This typically refers to criteria regarding tracking accuracy, adaptation speed, noise sensitivity, etc~\cite{gustafsson2000adaptive}. According to 
	\eqref{eq:sysCV}, the last three RCPs from the previous state are preserved during system propagation, while a new control point is appended for spline extension. Thus, the process noise covariance $\fQ$ is typically constructed with the following diagonal structure 
	\begin{equation}\label{eq:qmat}
		\fQ=(\omega\,\fI_{3d})\oplus(\nu\,\fI_d)\,,
	\end{equation}
	where usually a higher uncertainty is assigned to the newly added control point compared to the preserved ones, namely, $\omega\leq\nu$. During the update step, the latest RCP generally obtains a higher gain relative to the others. Over time, each RCP receives progressively less gain as it ages, until it is eventually removed from the state vector, and its estimate becomes fixed.
	
\begin{figure}[t]
		\centering
		\rotatebox[origin=c]{90}{$\bm{\tau=0.1}$}
		\begin{tabular}{ccc}
			$\bm{\omega/\nu=1}$ & $\bm{\omega/\nu=0.1}$ &$\bm{\omega/\nu=0.01}$\\
			\adjustbox{trim={0.23\width} {0.22\height} {0.19\width} {0.18\height},clip}{\includegraphics[width=0.22\textwidth]{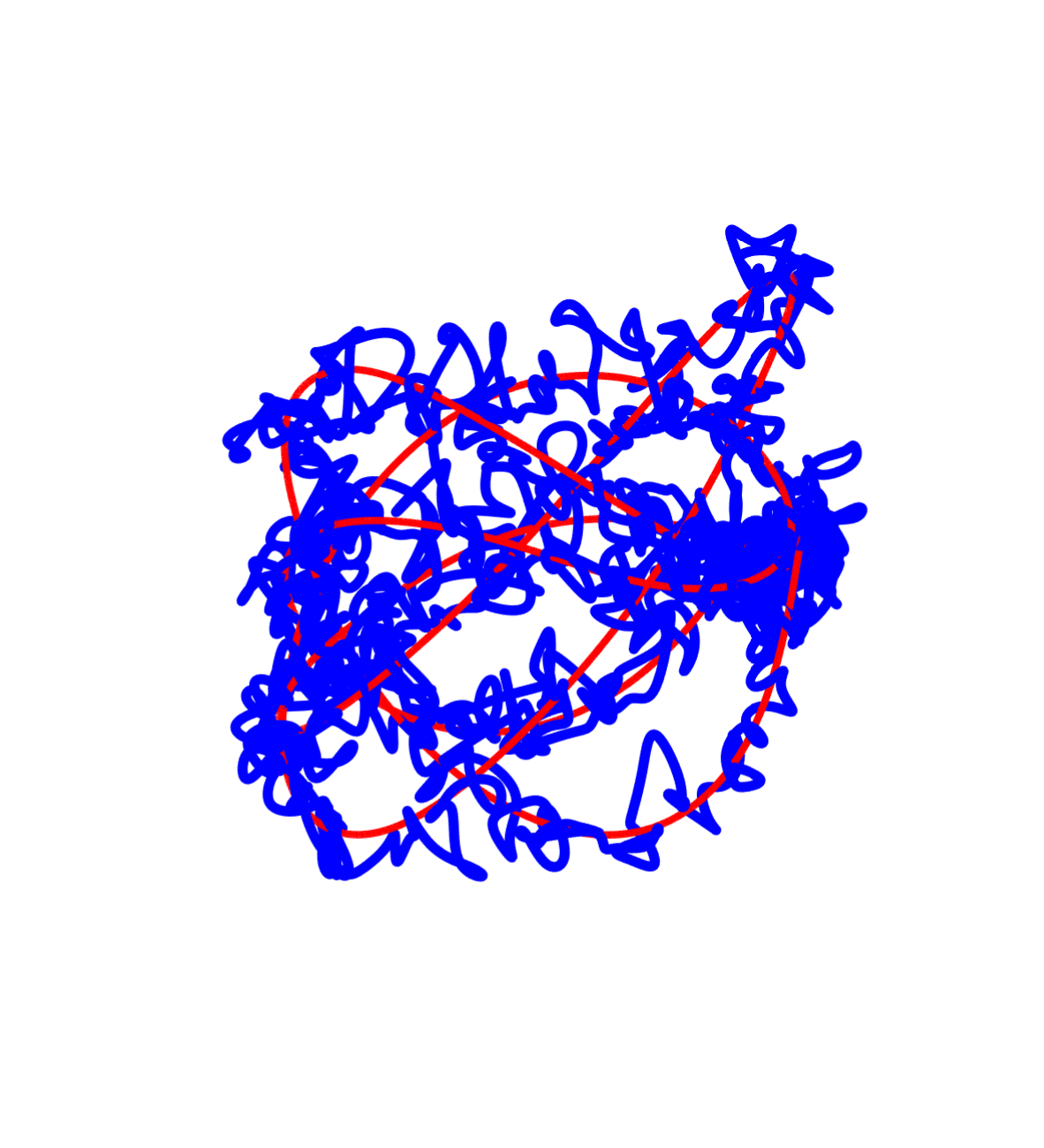}} &
			\adjustbox{trim={0.23\width} {0.22\height} {0.19\width} {0.18\height},clip}{\includegraphics[width=0.22\textwidth]{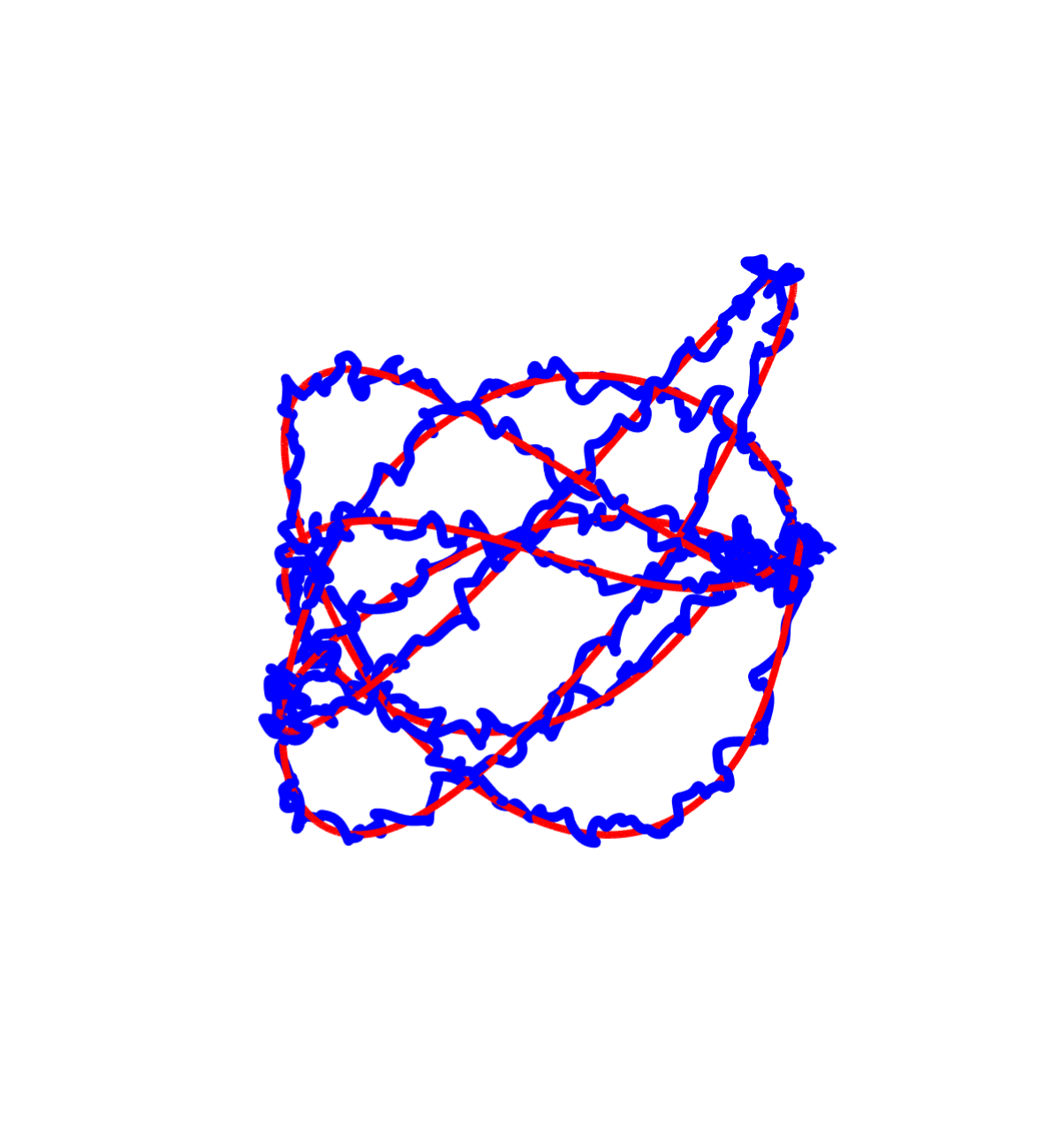}} &
			\adjustbox{trim={0.23\width} {0.22\height} {0.19\width} {0.18\height},clip}{\includegraphics[width=0.22\textwidth]{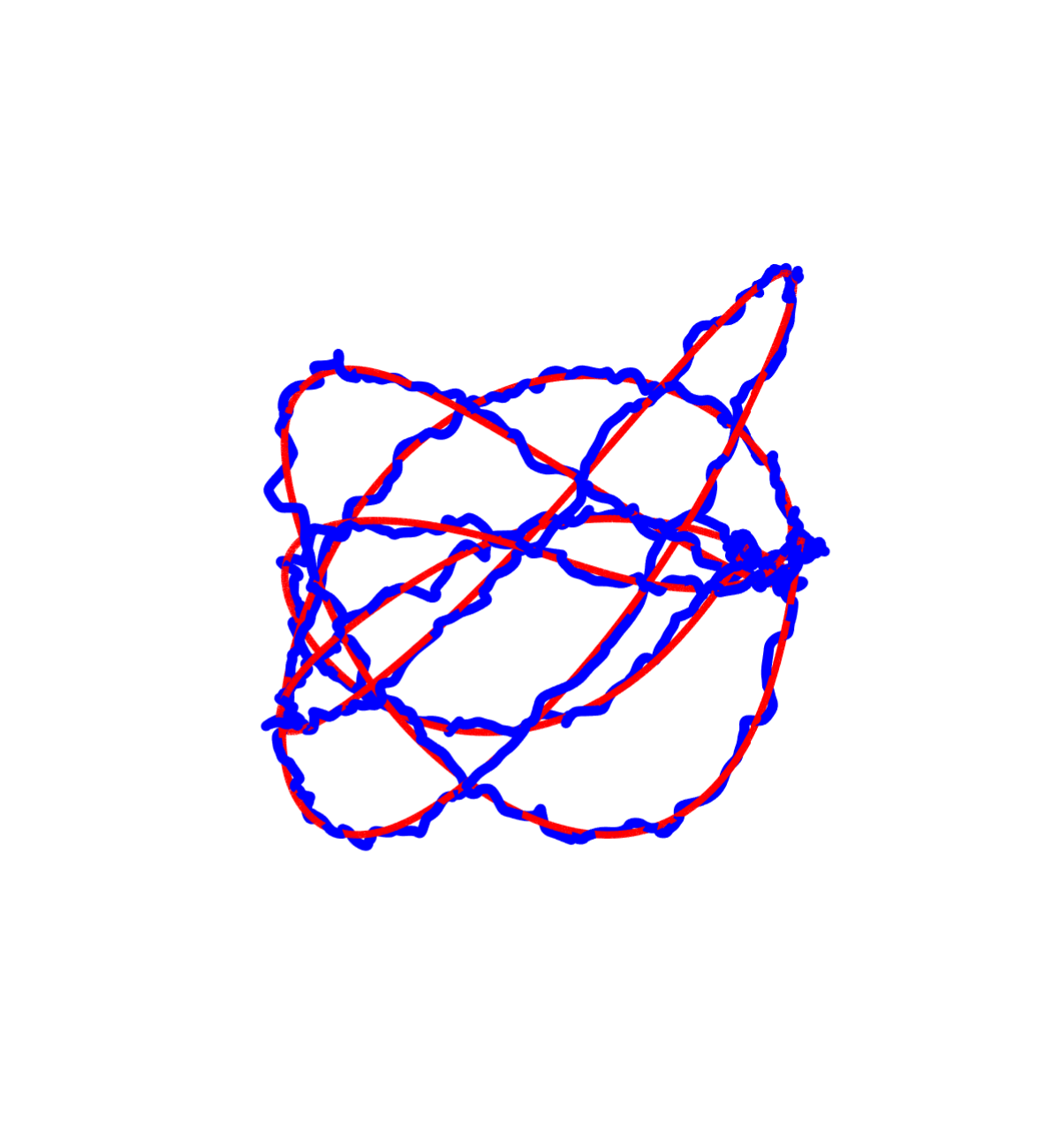}} \\
		\end{tabular}\\
		\rotatebox[origin=c]{90}{$\bm{\tau=1}$}
		\begin{tabular}{ccc}
			\adjustbox{trim={0.23\width} {0.22\height} {0.19\width} {0.18\height},clip}{\includegraphics[width=0.22\textwidth]{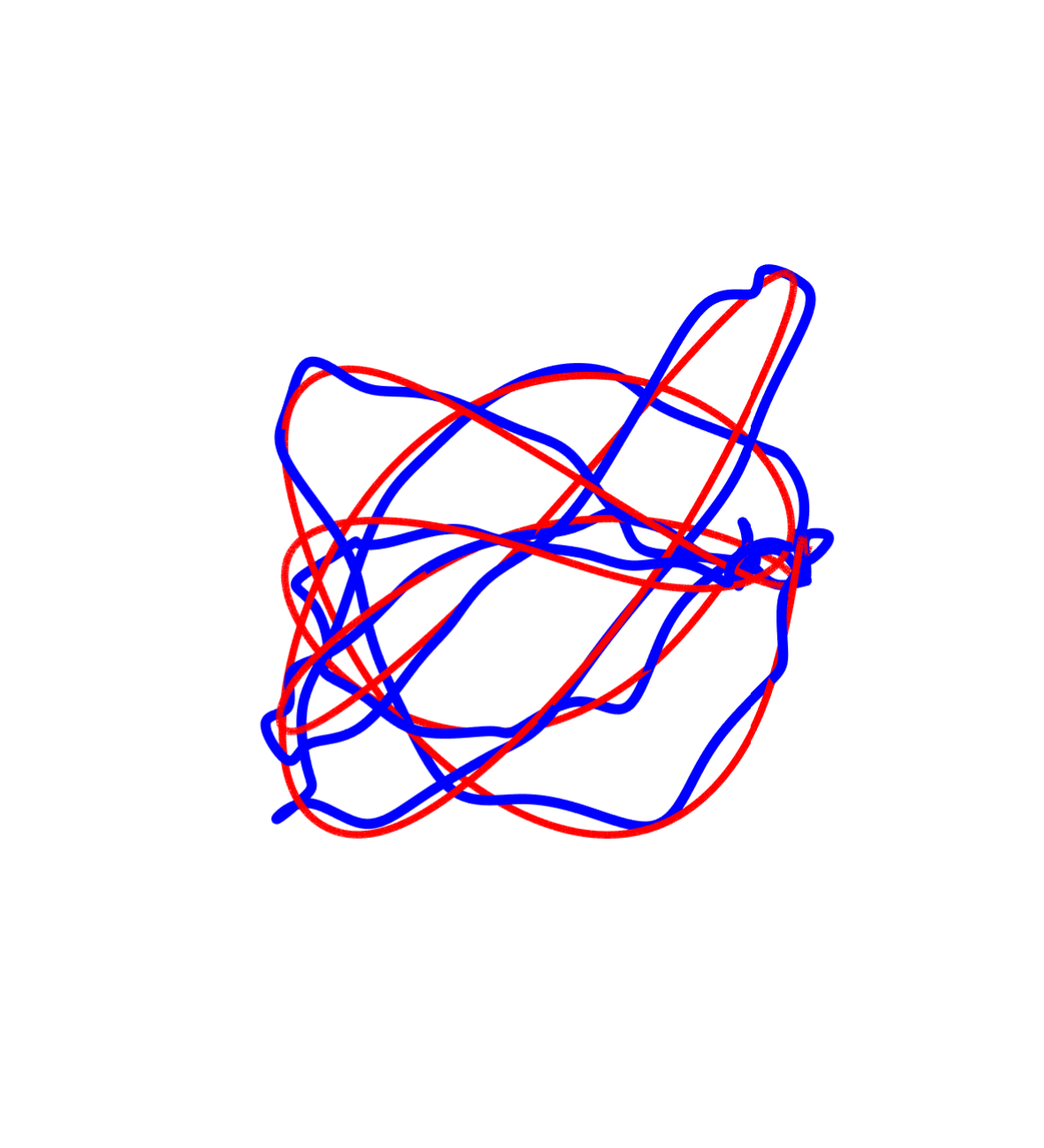}} &
			\adjustbox{trim={0.23\width} {0.22\height} {0.19\width} {0.18\height},clip}{\includegraphics[width=0.22\textwidth]{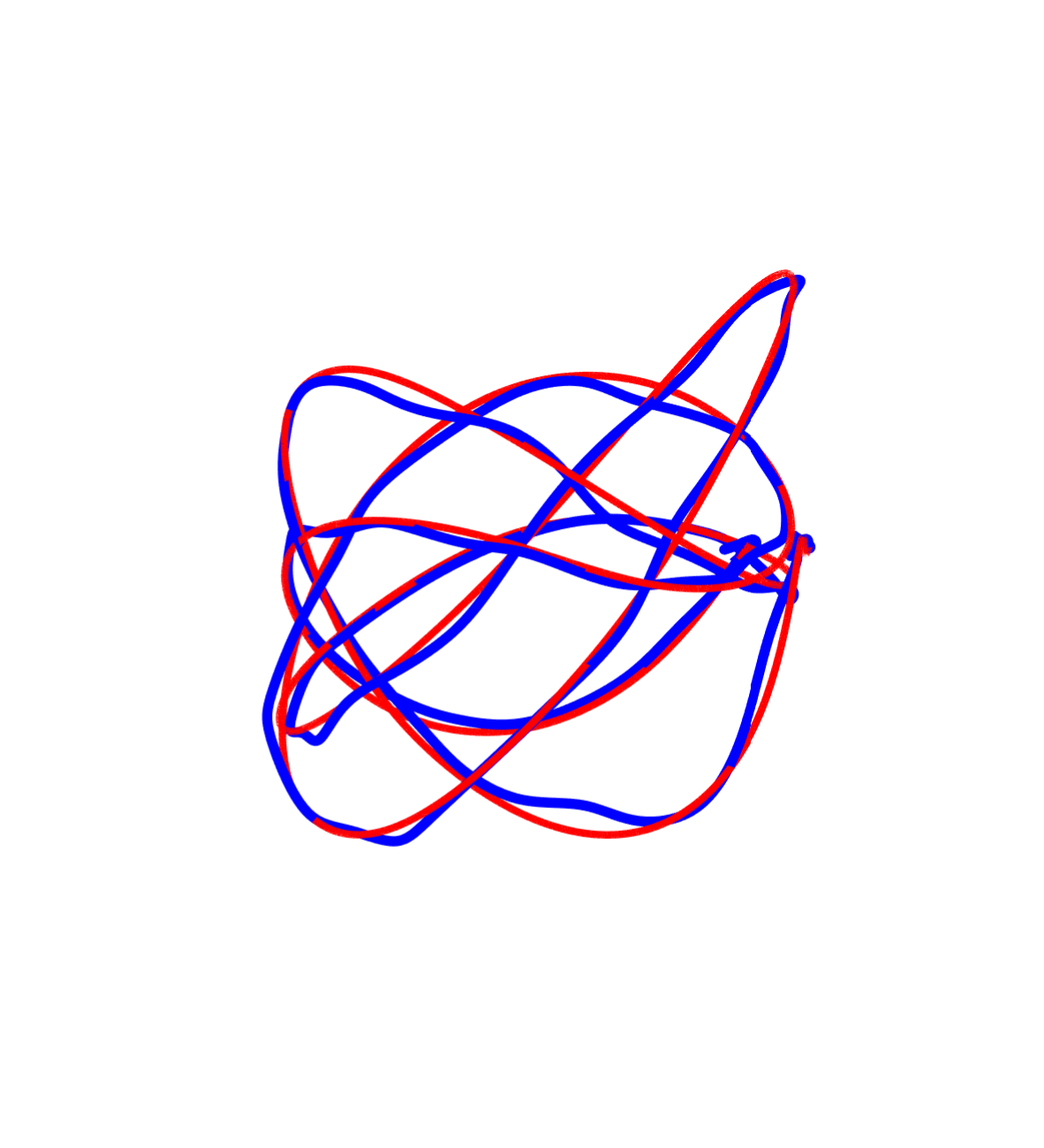}} &
			\adjustbox{trim={0.23\width} {0.22\height} {0.19\width} {0.18\height},clip}{\includegraphics[width=0.22\textwidth]{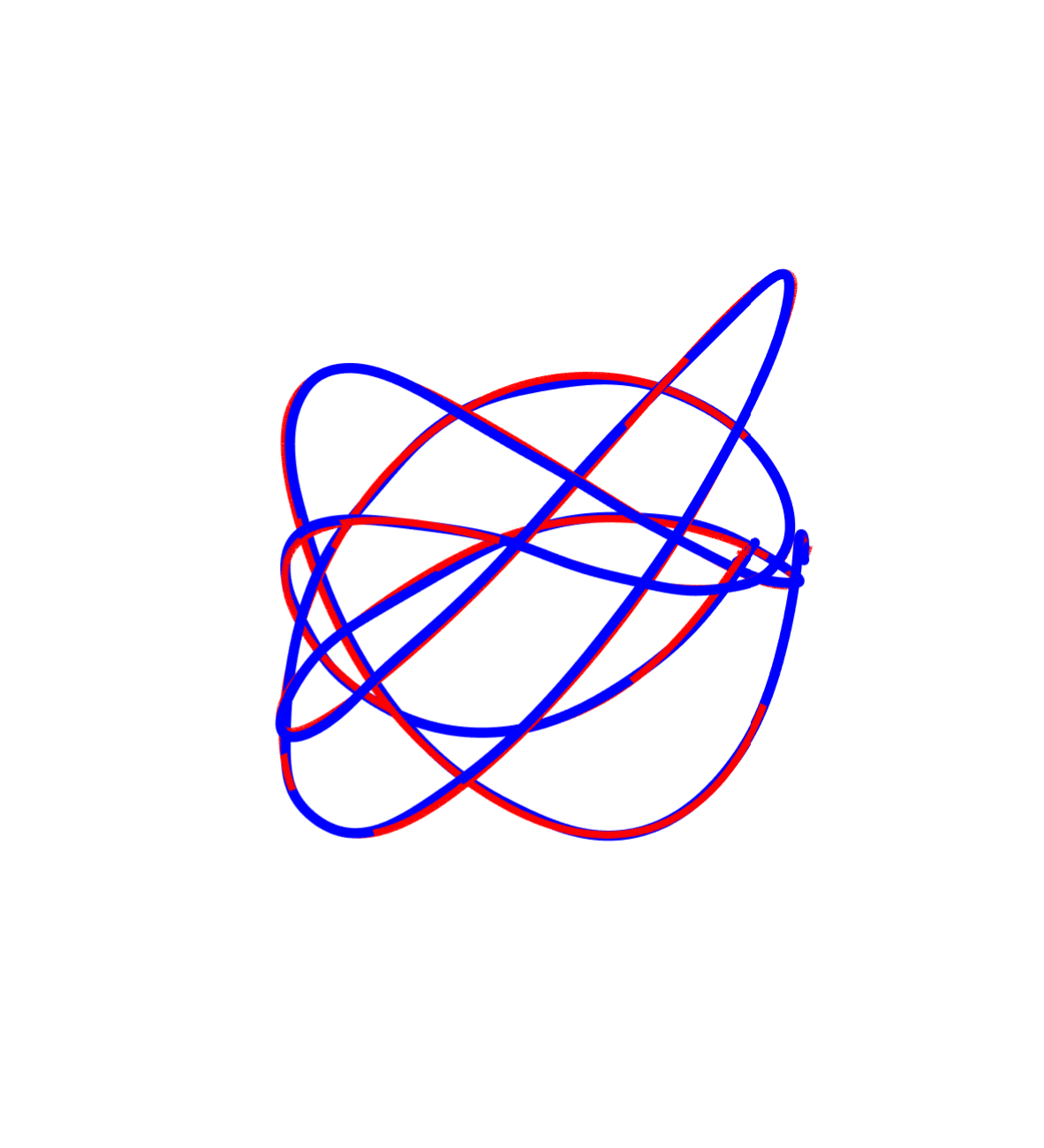}} \\
		\end{tabular}\\
		\rotatebox[origin=c]{90}{$\bm{\tau=6}$}
		\begin{tabular}{ccc}
			\adjustbox{trim={0.23\width} {0.22\height} {0.19\width} {0.18\height},clip}{\includegraphics[width=0.22\textwidth]{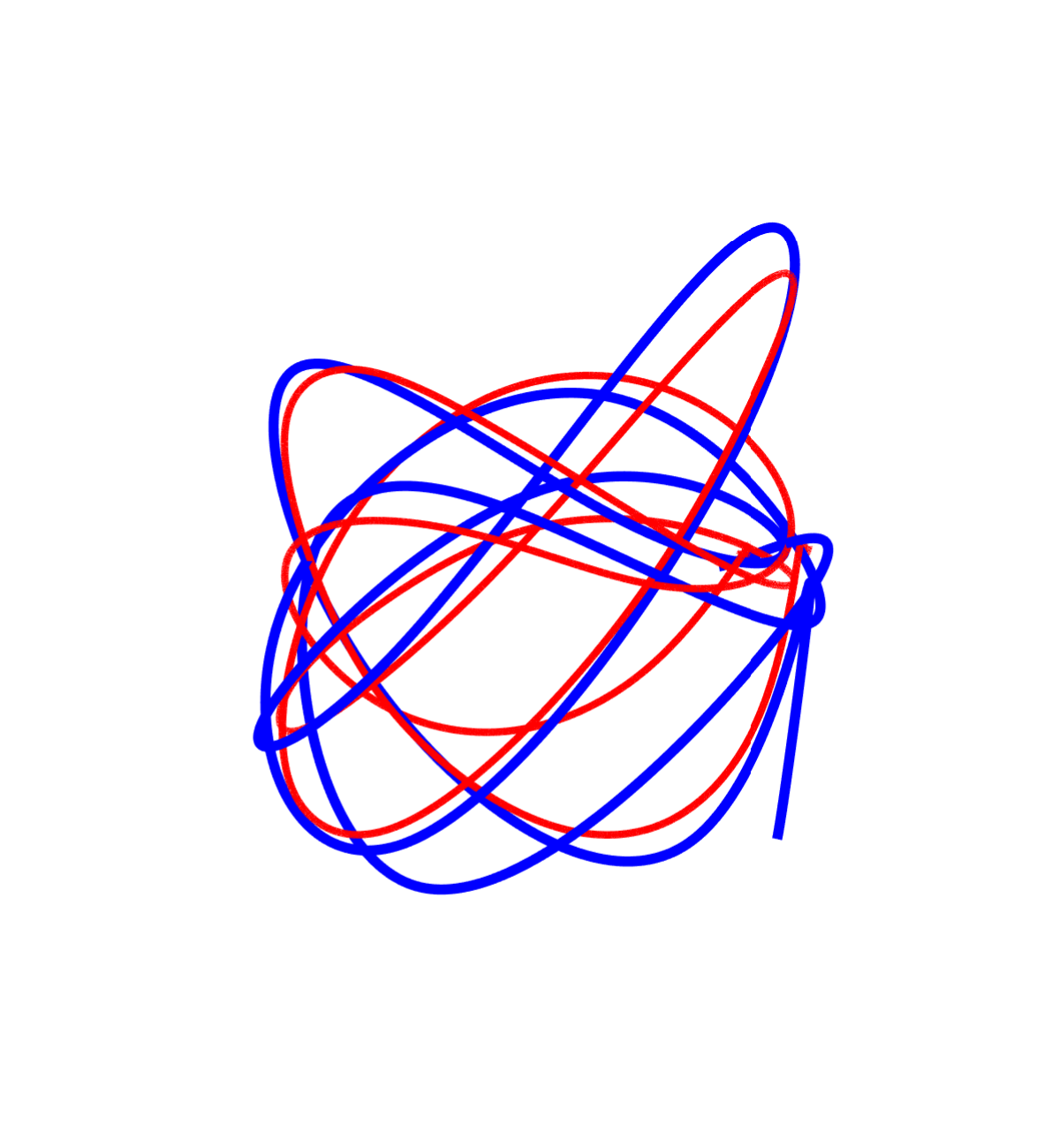}} &
			\adjustbox{trim={0.23\width} {0.22\height} {0.19\width} {0.18\height},clip}{\includegraphics[width=0.22\textwidth]{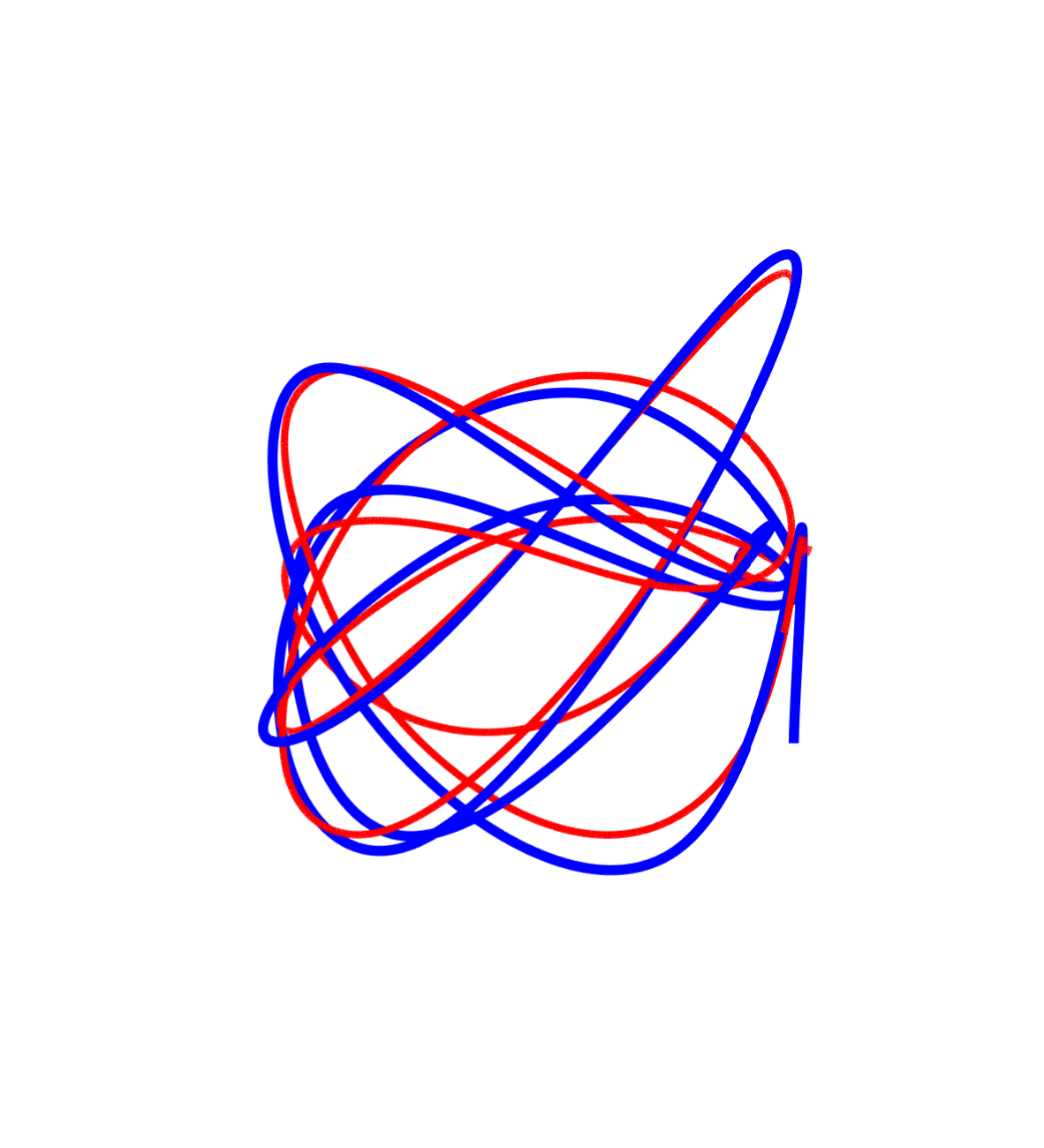}} &
			\adjustbox{trim={0.23\width} {0.22\height} {0.19\width} {0.18\height},clip}{\includegraphics[width=0.22\textwidth]{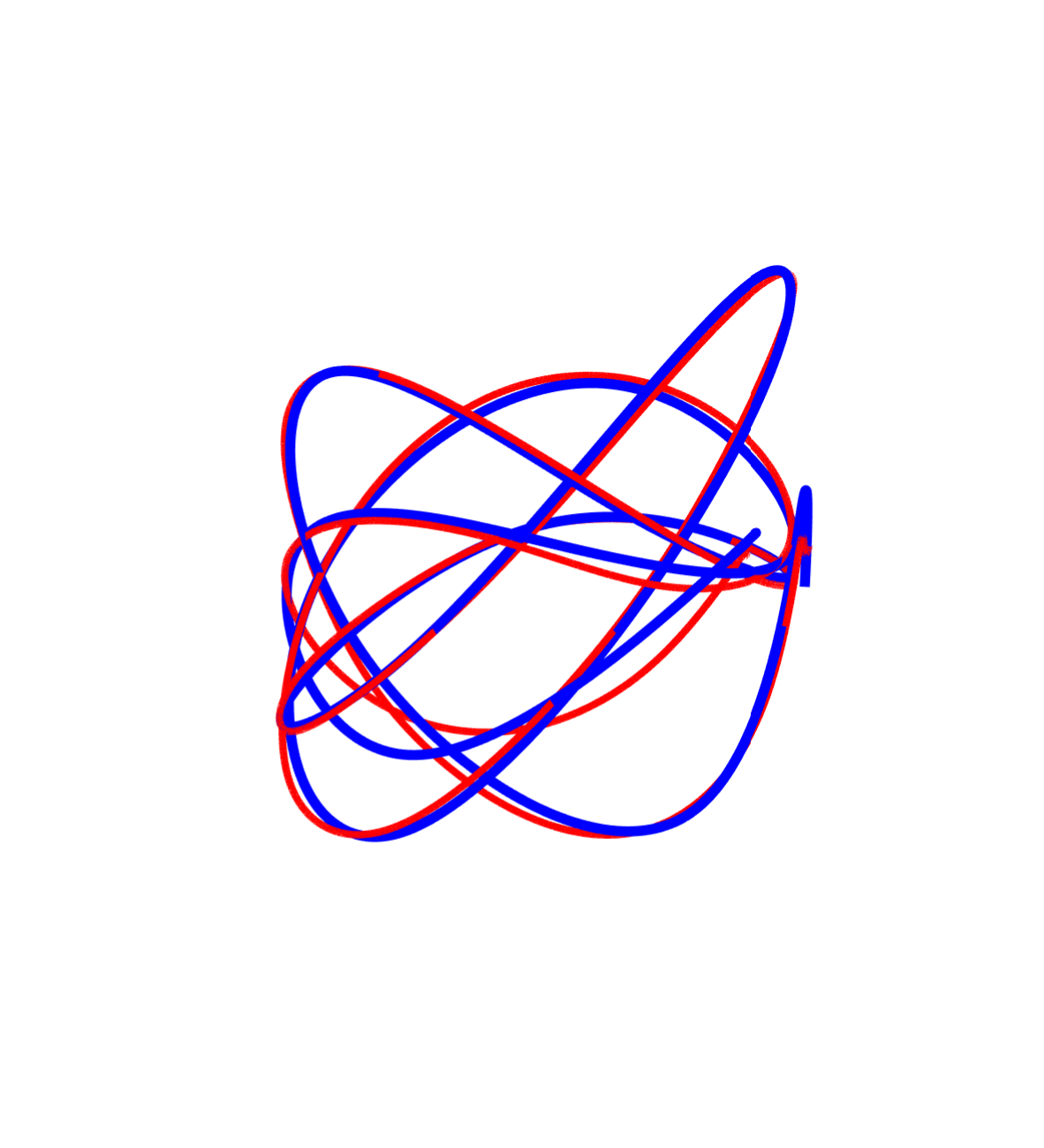}} \\
		\end{tabular}
		\caption{Trajectory estimates (blue) given by SERE \wrt ground truth (red) on \ttt{const1-trial2-tdoa2} in top view. We vary the temporal interval $\tau$ and noise ratios $\omega/\nu$ for RCPs propagation as discussed in Sec.\ref{sec:eva}-\ref{subsec:discuss}. The parameter set $(\omega/\nu,\tau)=(0.01,1)$ leads to the best tracking accuracy.}
		\label{fig:mat}
	\end{figure}

	To further justify the parameter tuning insights discussed earlier, we conduct a dedicated study using \ttt{const1-trial2-tdoa2} in \ttt{Syn-UTIL} under a higher noise level of $\text{R}^\toa=0.1$ (All other settings remain consistent with Sec.\ref{sec:eva}-\ref{subsec:synEva}). The process noise is configured according to \eqref{eq:qmat}, with $\omega\in\{0.1,0.01,0.001\}$ and $\nu=0.1$ governing the uncertainty of preserving and adding RCPs, respectively. Meanwhile, we vary the knot interval over $\tau\in\{0.1,1,6\}$. As shown in \figref{fig:mat}, given a fixed knot interval, reducing the relative uncertainty of preserving and adding RCPs (smaller $\omega/\nu$) leads to better tracking accuracy. Given the same process noise, coarser control points in time domain (larger $\tau$) induce less memory consumption for state storage, however, insufficiency in estimating complex trajectories. A small temporal interval $\tau$ can help mitigate this issue, however, may lead to overfitting under fixed sensing rates, producing overly dynamic motion estimates. In practice, tuning these parameters often comes to trade-offs among multiple factors, including motion dynamics, sensor data rate and hardware constraints, etc.
	
	\section{\uppercase{Conclusion}}\label{sec:conc}
	In this article, a principled study is presented on embedding B-splines into recursive Euclidean state estimation within the continuous-time domain. We conceptualize the recurrent control points (RCPs) and establish the spline-state-space (TriS) model based thereon by utilizing B-splines as a continuous-time intermediate in the standard state-space model, decoupling discrete-time state propagation from asynchronous observations. Further, the spline-embedded recursive estimation (SERE) scheme is proposed for continuous-time state estimation, including the probabilistic interpolation for dynamical state estimates. We conduct extensive evaluations using real-world and real-world-based synthetic sensor network scenarios for TDoA-only and ToA-inertial position tracking, respectively. Compared with conventional recursive filters, the SERE scheme demonstrates improved tracking accuracy and robustness, kinematic feasibility, storage consumption, and deployment flexibility.
		
	The proposed framework is fundamental and can be pragmatically deployed for real-world tasks\footnote{The original version of this paper was made public in July 2023.}, such as navigation, localization, and target tracking, as demonstrated in~\cite{xia2024,talbot2024continuous}. Significant potential remains for further enhancement. One promising direction is the development of dedicated methods for automatic parameter tuning adaptively to application scenarios. Another natural extension is to enable 6-DoF motion tracking by incorporating spline modeling of spatial rotations~\cite{RAL23_Li}. Moreover, B-splines using non-uniform knot interval could improve both the efficiency and fidelity of motion modeling and estimation. Additionally, applying SERE to a broader range of multi-sensor, real-world scenarios is compelling~\cite{radar2021,Fusion18_Li,lv2021clins}, which may, in turn, provide valuable insights for characterizing and refining the overall design of spline-based estimation frameworks.
	
	\section*{ACKNOWLEDGMENT}
	The author would like to thank Prof.~Fredrik Gustafsson from Linköping University for valuable feedback and Ziyu Cao for technical discussions.
	
	\bibsection*{REFERENCES}
	\bibliographystyle{IEEEtran.bst}
	\bibliography{bibExternal.bib}
	
%	\begin{IEEEbiography}[{\includegraphics[width=1in,height=1.25in,clip,keepaspectratio]{figs/li.png}}]{Kailai Li} (Member, IEEE) received the B.E. degree in mechanical engineering and automation in 2014 from Tsinghua University, Beijing, China and the M.Sc. degree in automation engineering in 2017 from RWTH Aachen University, Aachen, Germany, including a master's thesis completed at ETH Zurich, Zurich, Switzerland. In 2021, he received the Ph.D. degree in computer science from Karlsruhe Institute of Technology (KIT), Karlsruhe, Germany, with highest distinction.
%		
%	Since May 2024, he has been an Assistant Professor with the Bernoulli Institute for Mathematics, Computer Science and Artificial Intelligence, University of Groningen, The Netherlands. From 2022 to 2024, he was a postdoc with the Department of Electrical Engineering, Linköping University, Sweden. His research focuses on agile and trustworthy sensing, inference, learning and control for mobile robotics.
%		
%	Dr. Li is the recipient of the Jean-Pierre Le Cadre Best Paper Award (first runner-up) by the International Society for Information Fusion (ISIF) at FUSION 2022.
%	\end{IEEEbiography}
	
\end{document}